\documentclass[12pt]{article}

\usepackage[margin=1in]{geometry}
\usepackage{setspace}
\usepackage{microtype}
\usepackage{float}
\usepackage{graphicx}
\usepackage{tabularx}
\usepackage[font=small]{caption}
\usepackage{subcaption}
\usepackage{booktabs}
\usepackage{amsmath,amssymb}
\usepackage[table]{xcolor}
\usepackage{tcolorbox}
\tcbuselibrary{skins,breakable}
\usepackage{placeins}
\usepackage{multirow}
\usepackage{abstract}
\tcbset{
  taskprompt/.style={
    enhanced,
    colback=gray!4,
    colframe=gray!55,
    fonttitle=\bfseries\small,
    coltitle=black,
    boxrule=0.4pt,
    arc=2pt,
    left=4pt, right=4pt, top=3pt, bottom=3pt,
    before skip=0pt,
    after skip=0pt,
  },
  rubricbox/.style={
    enhanced,
    colback=blue!3,
    colframe=blue!40!black,
    fonttitle=\bfseries\small,
    coltitle=white,
    boxrule=0.4pt,
    arc=2pt,
    left=4pt, right=4pt, top=3pt, bottom=3pt,
    before skip=0pt,
    after skip=0pt,
  },
  outputbox gpt/.style={
    enhanced, breakable,
    colback=green!4,
    colframe=green!45!black,
    fonttitle=\bfseries\small,
    coltitle=white,
    boxrule=0.4pt,
    arc=2pt,
    left=4pt, right=4pt, top=3pt, bottom=3pt,
    before skip=0pt,
    after skip=0pt,
    title=#1,
  },
    outputbox ds/.style={
    enhanced, breakable,
    colback=cyan!6!white,
    colframe=cyan!60!blue,
    fonttitle=\bfseries\small,
    coltitle=white,
    boxrule=0.4pt,
    arc=2pt,
    left=4pt, right=4pt, top=3pt, bottom=3pt,
    before skip=0pt,
    after skip=0pt,
    title=#1,
  },
}

\usepackage[colorlinks=true, citecolor=blue, linkcolor=blue, urlcolor=blue]{hyperref}

\usepackage{parskip}

\usepackage[round,authoryear]{natbib}
\usepackage{titling}
\pretitle{\centering\LARGE\bfseries}
\posttitle{\par\vspace{0.6em}}
\preauthor{\centering\large}
\postauthor{\par\vspace{0.4em}}
\predate{\centering\normalsize}
\postdate{\par\vspace{0em}}

\usepackage{fancyhdr}
\title{\textsc{CentaurBench: Benchmarking LLM Capabilities on Augmenting vs. Automating Real-World Work Tasks}}

\author{%
  \begin{tabular}{c@{\hspace{4em}}c}
    Pattaraphon Kenny Wongchamcharoen\thanks{Corresponding author: \href{mailto:pattaraphon.kenny@berkeley.edu}{pattaraphon.kenny@berkeley.edu}. All authors are members of the Data
      Innovation and AI Lab (DIAL) at the Haas School of Business,
      UC Berkeley. The reproducibility code is available at \url{https://github.com/kennywong524/best-player-not-best-coach}. An interactive results dashboard is available at \url{https://kennywong524.github.io/centaur-benchmark/}} & Kris Gulati \\
    {\small UC Berkeley} & {\small UC Berkeley} \\[0.9em]
    Min Min Fong & Abhishek Nagaraj \\
    {\small UC Berkeley} & {\small UC Berkeley and NBER} \\
  \end{tabular}%
}

\date{July 2026\\[0.25em]\small Working Paper}

\begin{document}

\maketitle

\begin{abstract}
\footnotesize
Most LLM benchmarks rank models on their ability to automate work tasks. In practice, however, models are often used to assist other (human or LLM) agents. The question that drives model selection is therefore not only which model produces the best output, but which model most improves the work of another (weaker) agent. We introduce a unified framework that evaluates the capability of models to automate and augment another agent's performance. Across seven economically grounded real-world tasks, an assistant model writes assistance text for a standardized lower-capacity worker model, which produces the deliverable. In automation mode, the assistant produces the output directly. Outputs are scored through blind pairwise comparisons by an LLM judge panel with task-specific rubrics, replicated across ten runs. Rankings across the two regimes are only modestly correlated, and the automation winner loses augmentation on five of seven tasks. Assistance is not reliably positive. The unaided worker outranks every assisted condition on three tasks, and only one model's guidance beats no guidance on average.  These results suggest that automation ability is an incomplete proxy for assistance quality, motivating benchmarks that evaluate models according to the roles they play in human-AI and multi-agent systems.

\end{abstract}

\newpage

\section{Introduction}

Since the explosion in the adoption of large language models (LLMs), researchers have sought to measure the growth in capabilities across a wide variety of benchmarks. This work has moved from measuring capabilities on abstract knowledge or reasoning capabilities \citep{hendrycks2021, srivastava2023} to more real-world, economically valuable tasks \citep{gdpval2025, profbench2025, occubench2026}. These benchmarks provide an excellent indication of a model's capacity to automate work. However, in many real organizational and personal settings, models do not produce the output themselves. They are often assisting other agents, such as a human or another, perhaps weaker, LLM to complete the task. In such settings, models help workers plan an approach, decompose a task into steps, identify potential errors, and revise their work while leaving responsibility for the final deliverable with the worker. A model that excels at completing a task independently may not be the model that provides the most useful guidance to someone else. Put differently, as research with humans shows, the best (AI) player may not be the best coach \citep{hinds1999, camerer1989}.

Existing research does not provide a general way to compare these two capabilities. On the one hand, large-scale computer science benchmarks primarily rank models as autonomous task solvers, including increasingly realistic evaluations of professional and economically valuable work \citep{hendrycks2021, liang2023, gdpval2025, sopbench2025, occubench2026, profbench2025, jobbench2026}. On the other hand, field experiments by economists and management scholars examine whether AI augments human performance, but generally evaluate a single model in a particular domain and against a limited set of experimental conditions \citep{noy2023, peng2023, dellacqua2023, chen2024, doshi2024, brynjolfsson2025, ju2026}.

There is a line of work on multi-agent systems. In one stream, past work finds that a model's performance in isolation does not reliably predict its performance in interactive settings \citep{collins2024, chang2025, vaccaro2024}. Research has also studied how stronger models can teach weaker agents \citep{saha2023} or human learners \citep{macina2025}, and how AI can complement human decision makers \citep{bansal2021}. These studies evaluate assistance within a single domain (e.g. reasoning accuracy, tutoring pedagogy etc.), and score the assistance interaction itself. No existing benchmark compares models’ ability to assist and automate on an identical set of work-related tasks. Indeed, \citet{haupt2025} argue that model evaluation should move beyond an ``imitation game'' of autonomous task completion toward \emph{centaur evaluations} that measure how models augment human performance, yet no research operationalizes this call, jointly evaluating models as automators and as augmenters across multiple professional domains. Motivated by this gap, in this paper, we explore how models' automation capability correlates with its augmentation capability and explore how this correlation itself varies across tasks and models.

We evaluate LLMs as automators and augmenters across seven heterogeneous professional tasks. Our pipeline evaluates a set of frontier LLMs across two axes simultaneously: usage mode (automation versus augmentation) and task domain. In automation mode, each focal model completes a professional task end-to-end. In augmentation mode, each focal model provides assistance text to a fixed worker model, which then completes the task. Holding the worker constant isolates the value of the assistant's guidance from the executor's own capability, producing a direct measure of assistant models' quality rather than team outcome. This approach allows us to measure differences in augmentation and automation to any model-task pairing without a separate field experiment. Both outputs are evaluated by a panel of LLM judges using task-specific rubrics, with model families excluded from judging their own outputs (e.g. Claude Opus 4.8 cannot judge other Claude model outputs). The result is a rank matrix at the task and model level for both the augmentation and automation regimes.

Three key results are worth highlighting. First, we find that model rankings diverge meaningfully across usage modes and task domains. The strongest autonomous task solver is not necessarily the strongest assistant, and no single model dominates across augmentation tasks. The model-level rank correlation is about $0.48$ between augmentation and automation regimes and itself varies, ranging from $-0.04$ (travel planning) to $0.85$ (tax preparation) across tasks. Second, assistance is not always beneficial. While GPT-5-mini ranks first for both automation and augmentation schemes, the unaided worker baseline, in which GPT-3.5-Turbo completes the task without guidance from another model, achieves the second best average rank in the augmentation comparison. Guidance that is overly complex, poorly matched to the worker, or miscalibrated to the task can leave the downstream worker worse off than working alone. Third, reversals in model performance across regimes at the task level are large and stable. For example, the strongest automation model on our market trends analysis task, Claude-Opus-4.8 (mean rank 2.05), is among the weakest assistants on the same task (8.15), while GPT-4.1 shows the mirror-image profile on counseling (a lower-ranked direct solver but the strongest assistant).

Our study makes three contributions. First, we introduce a common evaluation framework for comparing models as both autonomous task solvers and assistants across professional domains. Second, by holding the downstream worker constant, we provide a direct measure of the marginal value of model-generated guidance and show that automation performance does not determine augmentation performance. Third, we demonstrate that the value of assistance depends on both the model and the task, challenging the use of a single general-purpose leaderboard for organizational model selection. These findings have direct implications for organizations choosing and deploying LLMs. The relevant question is not simply, ``Which model is best?'' but rather, ``Which model is best for this role and this task?'' A model selected to produce finished deliverables may differ from the model selected to support a worker through planning, critique, or revision. More broadly, our results suggest that future benchmarks should move beyond agent-alone performance and incorporate role-specific, workflow-aware evaluations of models as assistants, coaches, and collaborators.

\section{Related Literature}

Our study connects research on LLM benchmarking, field experiments on AI-assisted work, and emerging frameworks for evaluating human--AI complementarity.

\subsection{LLM Benchmarks and Autonomous Performance}

A large computer-science literature evaluates LLMs by asking them to complete tasks autonomously. Early general-purpose benchmarks emphasize knowledge and reasoning across broad collections of tasks. MMLU, for example, tests knowledge across 57 subjects \citep{hendrycks2021}, while HELM evaluates models across a wide range of scenarios and evaluation dimensions \citep{liang2023}. These benchmarks have played an important role in standardizing model comparisons, but their basic unit of evaluation remains the output produced by a model working independently.

More recent benchmarks have moved toward realistic professional and economically grounded tasks. These evaluations test models on industrial workflows \citep{sopbench2025}, occupational tasks spanning specialized domains \citep{occubench2026}, and work products designed to approximate economically valuable professional activity \citep{gdpval2025, profbench2025}. JobBench further makes task selection more worker-centered by constructing 130 agentic tasks across 35 occupations from duties that workers report wanting to delegate, and by evaluating the resulting work against fact-anchored criteria \citep{jobbench2026}. This represents an important shift from selecting tasks solely according to economic value or technical feasibility toward also considering which activities workers want AI systems to perform. This progression has increased the realism and economic relevance of model evaluation. Nevertheless, operationally, these benchmarks still evaluate the model as the party responsible for completing the selected task and producing the final deliverable end-to-end. They therefore provide increasingly realistic measures of automation, but do not directly measure the value a model creates by augmentation.

That focus leaves a distinct capability undermeasured. In many workplace applications, the model does not directly produce the final deliverable. Instead, it helps another party plan, execute, review, or revise the work. Evidence that autonomous performance does not necessarily predict interactive performance reinforces this concern \citep{collins2024, chang2025}. A benchmark that identifies the best independent task solver may therefore provide limited guidance about which model should be deployed as an assistant.

\subsection{AI Augmentation and Workplace Productivity}

A parallel literature in economics and management examines how access to generative AI affects worker performance. Across several settings, experiments have found substantial productivity effects. ChatGPT reduced the time required for professional writing tasks while improving output quality \citep{noy2023}; GitHub Copilot accelerated completion of a programming task \citep{peng2023}; and a large-scale deployment of a generative-AI assistant increased productivity among customer-support workers \citep{brynjolfsson2025}. Related studies examine consultants \citep{dellacqua2023}, advertising professionals \citep{chen2024, ju2026}, and creative writers \citep{doshi2024}.

These studies establish that generative AI can affect worker productivity and that its effects depend on the task, worker, and mode of interaction. At the same time, field experiments are necessarily narrow by design. They typically study one model or tool, one population, and one domain, often relative to an unaided control condition. Their purpose is generally to estimate whether access to a particular AI system improves performance in a particular setting, rather than to compare which of several models would provide the greatest assistance to the same worker.

The evidence also cautions against treating human--AI collaboration as automatically beneficial. In a meta-analysis of 106 experiments, \citet{vaccaro2024} find that human--AI combinations frequently underperform the better of the human-alone and AI-alone conditions. The value of assistance may depend on how capabilities are distributed between the parties, how work is allocated, and whether the system's guidance is appropriately calibrated to the worker. These findings motivate evaluations that compare not only whether assistance is available, but also which model provides that assistance and how its value varies across tasks.

This question is economically consequential because automation and augmentation can produce different organizational and distributional effects. \citet{agrawal2023} argue that automating one worker's tasks can create augmentation opportunities elsewhere in a production process, while \citet{marguerit2025} distinguish the labor-market effects of augmentation-oriented and automation-oriented AI. Worker preferences may also favor collaborative arrangements: a nationwide survey of 1,500 U.S. workers across 104 occupations found that ``equal partnership'' with AI, rather than replacement, was the dominant preference in nearly half of the occupational categories studied \citep{shao2025}. Together, this research highlights the importance of evaluating AI systems in the roles through which they are actually deployed.

\subsection{Automation, Augmentation, and Human--AI Complementarity}

A growing body of work more directly distinguishes automation from augmentation. \citet{haupt2025} argue that evaluation by imitation privileges systems that reproduce human outputs and replace human effort, rather than systems designed to improve and augment human performance. They call for evaluations in which humans and AI systems solve tasks together. This critique shifts the object of evaluation from the model's standalone capability to the value it creates within a larger workflow.

\citet{fugener2026} formalize the distinction between automation and augmentation, showing that the optimal allocation of work depends on the structure of complementarity between human and AI capabilities. A system that performs well independently need not create the greatest value when paired with another decision-maker or worker. Their analysis therefore provides a theoretical basis for expecting model rankings to vary across autonomous and assistive roles. A growing set of efforts evaluates this type of assistance empirically, but each within a narrow scope. Upwork's Human+Agent Productivity Index (HAPI) evaluates frontier agents on approximately 300 paid client projects, and finds that expert involvement can substantially improve agent performance \citep{upwork_hapi2025}. However, a human expert assists the AI, reviewing and correcting the agent's output, so the human occupies the coaching role and the model is the party being helped. MathTutorBench benchmarks the pedagogical quality of LLM tutors, finding that stronger problem-solving ability does not automatically imply better teaching \citep{macina2025}. \citet{gandhi2025} study strong--weak model collaboration for code generation, in which an expert model supplies plans for an inferior model to execute, reducing cost while preserving performance. \citet{saha2023} examine a teacher -- student simulation between two LLMs, showing that the stronger model's explanations improve the weaker model's performance when explanations are personalized to the student. Each of these studies is confined to a single domain (e.g. tutoring, coding, freelance projects) and to a single human - LLM or model - model pairing, rather than comparing many models against one another in the assistant role.

Our approach differs from these studies by spanning a range of real-world professional tasks and placing models into an assistance/guidance role to a fixed worker. Importantly, the latter allows measurement of the marginal value of an assistant model's guidance rather than its joint outcome as a team. Rather than asking only whether automation or augmentation is optimal in a particular setting, we compare the same set of models under both modes and across multiple professional domains. This design allows us to examine whether relative model rankings are stable across roles and whether a model's autonomous capability is also an indication of the value of its guidance to a fixed downstream worker. In sum, our study operationalizes this distinction as a cross-model, cross-mode, and cross-task empirical evaluation.

\section{Methodology}
\label{sec:methodology}

    \subsection{Overview}

\begin{figure}[H]
\centering
\includegraphics[width=\linewidth]{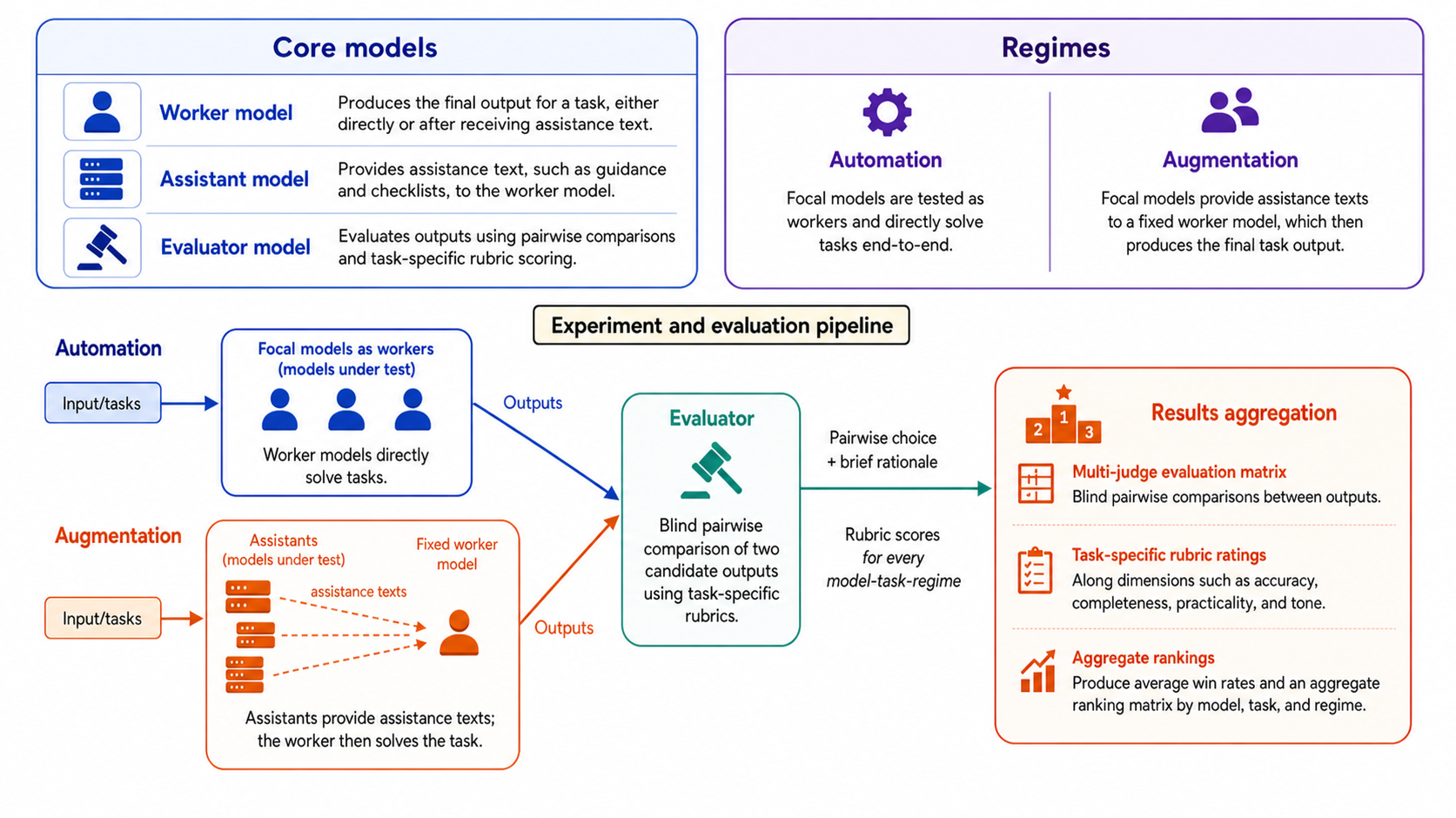}
\caption{\textbf{Methodology pipeline.} The framework evaluates models in two usage modes: automation, where focal models directly solve tasks, and augmentation, where focal models provide an assistance text to a fixed worker model. Outputs are evaluated through rubric-guided pairwise comparisons and aggregated into model rankings by task and usage mode.}
\label{fig:methodology-pipeline}
\end{figure}

We introduce a framework that evaluates large language models not only as direct task solvers but as assistants that help another agent do the work. We study two usage modes. In the \emph{automation} mode, as in most existing benchmarks, a focal model completes a professional task end-to-end. In the \emph{augmentation} mode, a fixed lower-capability worker
model completes the task with help from an assistant model: the worker requests assistance, the assistant model replies with an assistance text - a plan, checklists, constraints, and self-review steps - and the worker uses that support to produce the final deliverable. The assistant model never writes the deliverable itself, so what varies across augmentation trials is the quality of its guidance, not whether it simply did the work. We use GPT-3.5-Turbo as the
worker for two reasons: it approximates worker performance on our pilot tasks, and it represents the cheaper, lower-capability models that are often assigned the execution role in multi-agent settings. Figure~\ref{fig:methodology-pipeline} summarizes the full workflow.

Every output is evaluated in both usage modes through blind pairwise comparisons by a panel of four LLM judges using task-specific and general rubrics. Each judge is barred from scoring outputs produced by its own model family (leave-one-family-out masking) to limit self-preference, and we replicate the full pipeline across ten independent runs. The result is a cross-mode, cross-task rank matrix over ten models: Claude-Opus-4.8, Claude-Sonnet-4.6, Gemini-3.1-Pro, DeepSeek-V3.1, GPT-5-Mini, GPT-OSS-120B, GPT-O4-Mini, GPT-4.1, GPT-O3-Mini, and GPT-3.5-Turbo. For each model, we can ask not only whether it ranks higher as a direct solver or as an assistant, but also whether that ordering holds across tasks. The framework therefore produces a profile for each model rather than a single score. A model may complete structured execution tasks well on its own yet be a weaker assistant on open-ended, human-facing tasks; another may rank lower as a direct solver but produce assistance text that clearly improve counseling, tutoring, or planning outputs. Capturing this split is our core methodological
contribution.

\subsection{Task Selection and Prompt Design}

We construct a benchmark of seven tasks representing activities commonly encountered in everyday and professional settings and associated with economically meaningful forms of work: counseling, market analysis, meal planning, operations research, tax preparation, travel planning, and tutoring. Tasks were chosen to vary in structure, domain knowledge, risk, and the degree to which human-facing judgment matters. Each task specifies an observable deliverable with concrete requirements, allowing evaluation to assess whether outputs satisfy stated constraints objectively rather than relying on subjective impressions of quality. Figure~\ref{fig:tasks} presents the full task set. The assistant model set and judge panel are reported in Appendix~\ref{app:models}.

To make evaluation reliable, each task prompt is paired with a micro-rubric. The prompt specifies the required deliverable and the components that must be present. The rubric mirrors those components and scores each dimension on a 1--10 scale. Each task includes task-specific dimensions, such as dietary safety for meal planning or mathematical correctness for tutoring, as well as general dimensions such as instruction following, accuracy, practical usefulness, organization, and tone. We evaluate models pairwise to enable contrastive grading (which we found worked better), where the judge can compare models to determine winners rather than grade inherently subjective outputs. However, we also develop the task-specific rubrics to help ground the judge's decision, and also as a way for humans to reason about why certain outputs were judged better than others.

\begin{figure}[H]
\centering
\renewcommand{\arraystretch}{1.15}
\setlength{\tabcolsep}{6pt}
\small

\definecolor{planyellow}{RGB}{255,247,222}
\definecolor{profgreen}{RGB}{231,243,231}
\definecolor{humanpurple}{RGB}{237,236,250}
\definecolor{headnavy}{RGB}{16,42,94}

\begin{tabularx}{\linewidth}{
    >{\raggedright\arraybackslash\hsize=0.85\hsize}X
    >{\raggedright\arraybackslash\hsize=1.05\hsize}X
    >{\raggedright\arraybackslash\hsize=1.00\hsize}X
    >{\raggedright\arraybackslash\hsize=1.10\hsize}X}
\toprule
\rowcolor{headnavy}
\textcolor{white}{\textbf{Task}} &
\textcolor{white}{\textbf{Source}} &
\textcolor{white}{\textbf{Task type}} &
\textcolor{white}{\textbf{Deliverable}} \\
\midrule
\rowcolor{planyellow}
Menu Planning & GDPval & Structured planning & Seven-day constrained meal plan \\
\rowcolor{planyellow}
Travel Planning & GDPval & Structured planning & Tokyo itinerary under fixed budget \\
\rowcolor{profgreen}
Market Trends Analysis & GDPval & Professional / analytical & Client brief on natural gas trends \\
\rowcolor{profgreen}
Operations Research Analysis & Anthropic Economic Index & Professional / analytical & Executive logistics optimization memo \\
\rowcolor{profgreen}
Tax Preparation & Designed task & Professional / analytical & Tax discrepancy review and correction guidance \\
\rowcolor{humanpurple}
Counseling & Anthropic Economic Index & Human-facing interactive & Supportive counseling-style response \\
\rowcolor{humanpurple}
Tutoring Session Planning & Anthropic Economic Index & Human-facing interactive & Grade 3 lesson segment \\
\bottomrule
\end{tabularx}

\caption{\textbf{Tasks included in the benchmark.}
    The figure reports each task's source, broad task type, and required deliverable. Colors distinguish structured-planning tasks (yellow), professional or analytical tasks (green), and human-facing interactive tasks (purple). Tasks were drawn from GDPval and the Anthropic Economic Index, except for tax preparation, which was designed by the authors to provide a rule-based professional task with verifiable discrepancies. Full task prompts and evaluation rubrics are provided in Appendix~\ref{app:prompts}.}
\label{fig:tasks}
\end{figure}

\begin{figure}[H]
\centering
\includegraphics[width=\linewidth]{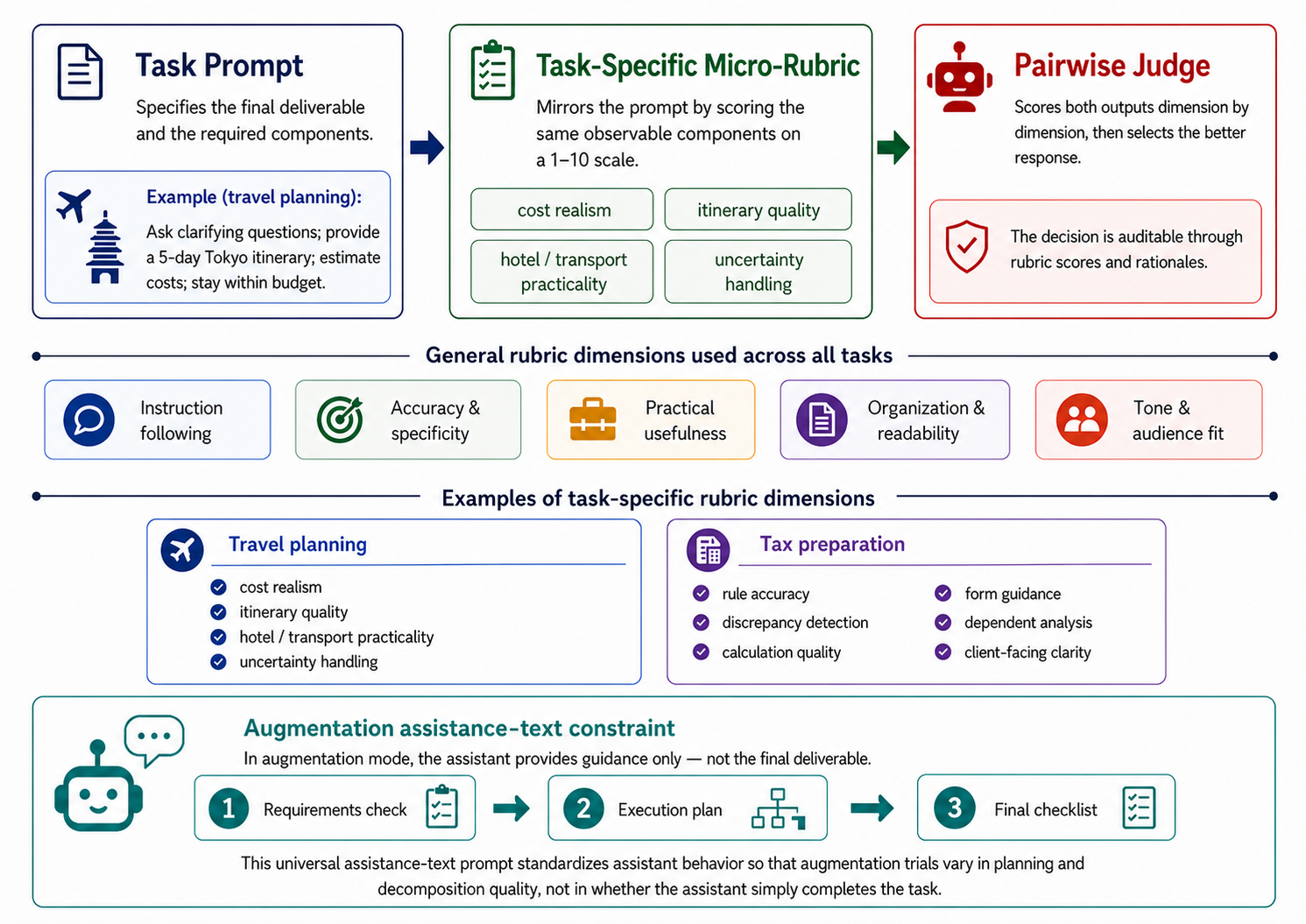}
\caption{\textbf{Prompt and rubric design.} Each task prompt specifies observable deliverable requirements, which are mirrored by task-specific micro-rubrics. General rubric dimensions are held constant across tasks, while task-specific dimensions vary by domain. In augmentation mode, the assistance text creation prompt constrains assistant models to provide process guidance rather than the final deliverable.}
\label{fig:prompt-rubric-design}
\end{figure}

This alignment between prompt and rubric is central to the framework. We do not ask evaluators to make an undifferentiated and subjective judgment of quality. Instead, they judge whether the response satisfies observable dimensions of the task that are quantifiable and comparatively objective to assess. For example, the travel-planning rubric separately measures cost realism, itinerary quality, hotel and transportation practicality, and handling of uncertainty. Meanwhile, the tax-preparation rubric separately measures rule accuracy, discrepancy detection, calculation quality, form guidance, dependent analysis, and client-facing clarity. Figure~\ref{fig:prompt-rubric-design} summarizes the alignment between task prompts, task-specific micro-rubrics, general evaluation dimensions, and the augmentation assistance text constraint.

Figure~\ref{fig:prompt-rubric-example} illustrates the prompt and rubrics for travel planning. Full prompts, assistance texts, and rubrics for all tasks appear in Appendix~\ref{app:prompts}. In augmentation mode, each assistant model receives a universal assistance text creation prompt that instructs it to produce a process-oriented plan for the worker, rather than completing the task itself. The prompt specifies a three-phase structure: a requirements check, an execution plan, and a final checklist. It additionally constrains the assistant to provide guidance only, explicitly prohibiting production of the final deliverable. This design ensures that what varies across augmentation trials is the quality of the assistant's planning and decomposition, not whether the assistant simply solved the task on the worker's behalf.

\begin{figure}[H]
\centering
\input{tables/table_prompts}
\caption{\textbf{Illustrative task prompt, evaluation rubric, and assistance-text
instruction.} The travel-planning example shows how observable task requirements (top) map onto task-specific rubric dimensions (middle). In augmentation mode, the assistant additionally receives the universal assistance text creation prompt shown above (bottom), which requires process-oriented guidance while prohibiting completion of the final deliverable. The complete prompts and rubrics are provided in Appendix~\ref{app:prompts}.}
\label{fig:prompt-rubric-example}
\end{figure}

Augmentation can be operationalized in many ways, ranging from an assistant that drafts content for the worker to edit, to one that supplies worked examples or directly corrects worker errors. Education research draws a similar distinction: procedural, strategic, and metagonitive scaffolding that shapes \emph{how} a learner approaches a task, versus conceptual scaffolding that supplies domain content \citep{hannafin1999}. We study process-oriented scaffolding, in which the assistant shapes how the worker approaches the task but never supplies solution content. This mirrors the educational ideal of teaching a learner to perform a task rather than performing it for them \citep{wood1976, vandepol2010} and isolates the value of guidance from the assistant's own task-solving ability. By design, it is a deliberately conservative form of assistance and lower bound on what augmentation can achieve, since it withholds conceptual, content-bearing support. Thus, the results should be read as characterizing this form rather than augmentation at large. However, the same fixed-worker methodology extends directly to richer forms of assistance, and mapping how augmentation value varies across richer types is a natural next steps our framework is designed to support. The full assistance prompt is provided in Appendix~\ref{app:prompts}.

\subsection{Evaluation Procedure}

\begin{figure}[H]
\centering
\includegraphics[width=\linewidth]{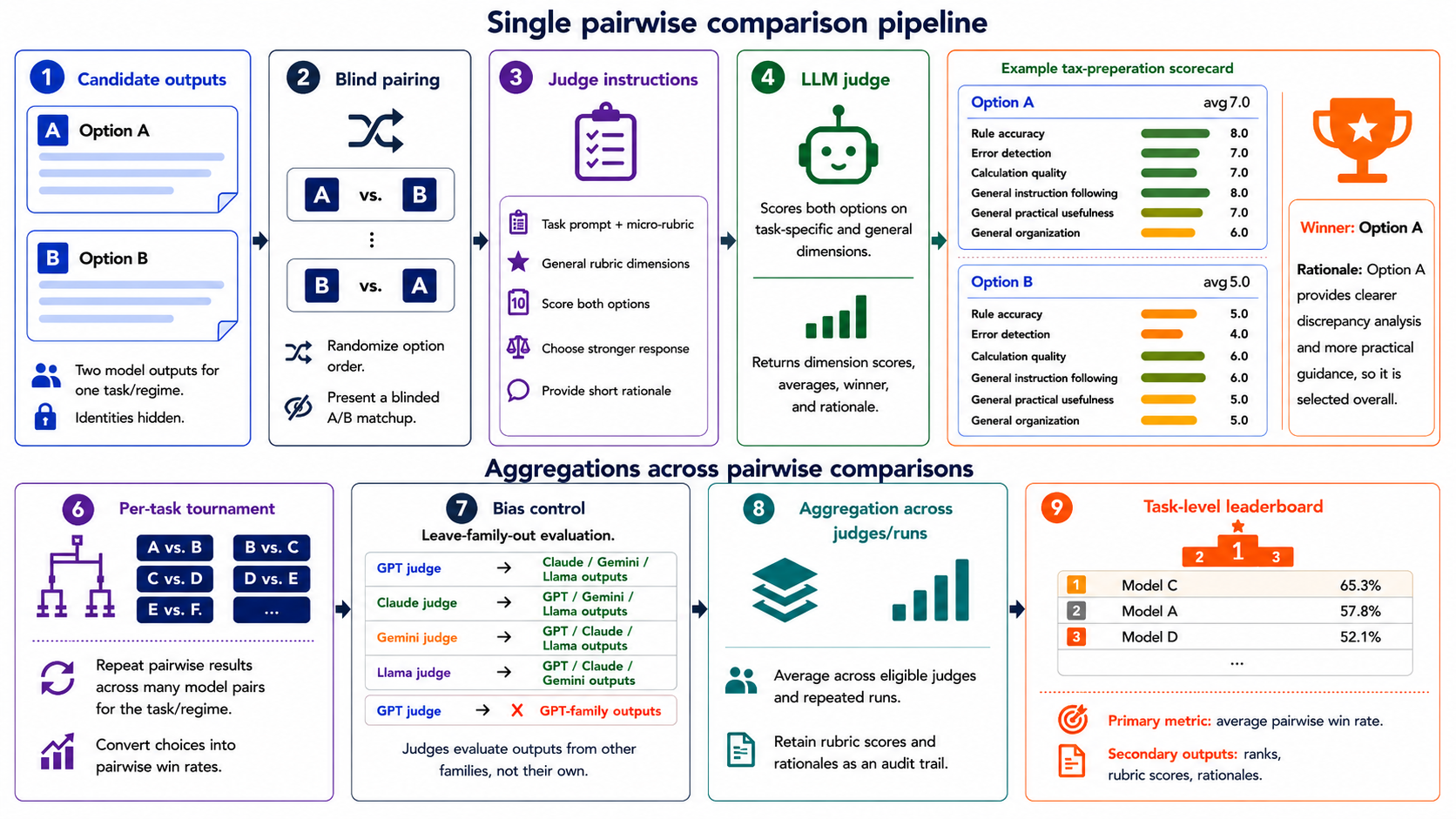}
\caption{\textbf{Rubric-guided pairwise evaluation pipeline.} Candidate outputs are anonymized, randomly paired, and evaluated by LLM judges using task-specific and general rubric dimensions. Pairwise choices are converted into win rates, aggregated across eligible judges and repeated runs, and reported alongside rubric scores and rationales as an audit trail.}
\label{fig:eval-pipeline}
\end{figure}

Following prior work on LLM-as-judge evaluation and pairwise preference benchmarking \citep{zheng2023,chiang2024}, we evaluate final outputs through blind pairwise comparisons conducted by a panel of four LLM judges. Figure~\ref{fig:eval-pipeline} summarizes the procedure.

For each task, usage mode, and independent run, we conduct a complete round-robin tournament among the ten candidate outputs. In augmentation, this candidate set includes the \emph{plain} baseline: the fixed worker operating without an assistance text. This baseline indicates whether receiving assistance improves the worker's output at all. Automation and augmentation
outputs are evaluated in separate tournaments and are never compared directly.

For every comparison, model identities are concealed and option order is randomized. The judge scores both responses from 1 to 10 on the task-specific and general rubric dimensions, computes an average score for each response, selects the stronger response, and provides a short rationale. We apply \emph{leave-family-out} evaluation: a judge does not evaluate comparisons containing outputs from its own model family. The model and judge panels are reported in Appendix~\ref{app:models}.

We treat pairwise comparison and rubric-based scoring as complementary components of our evaluation pipeline. Pairwise judgments provide the primary ranking signal because relative comparisons can recover expert-intended quality orderings more reliably and with lower annotation time than absolute rubric scores in judgment-intensive professional tasks \citep{yang2026judgmentbench}. More broadly, rubric-guided pairwise LLM judgments have been shown to align closely with human preferences \citep{zheng2023}. Pairwise choices alone, however, reveal which response is preferred without fully explaining why. Dimension-level rubric scores provide an inspectable diagnostic record, identifying the specific strengths and weaknesses underlying each comparison and supporting our qualitative analysis. Because rubric-based evaluation is itself sensitive to score calibration, scale use, and score-option ordering \citep{xu2026pointwisepairwise}, we use standardized rubric-derived rankings as a robustness check rather than as a replacement for pairwise comparison. We therefore use pairwise outcomes to construct the primary rankings and standardized rubric scores to interpret those rankings and test their robustness. Together, the two signals make the evaluation more reliable, transparent, and diagnostically useful. Across non-tied comparisons, the selected response receives the higher mean rubric score in 99.7\% of cases. In addition to the consistency between a judge's choice and its own rubric scores, we also measure agreement across judges. We separately assess inter-judge
reliability across all runs. Across 6,265 comparisons evaluated by at least two eligible judges, the judges select the same winner in 71.0\% of cases. Agreement is higher in automation (74.5\%) than in augmentation (67.8\%). Appendix~\ref{app:llm-judge-validity} reports the full LLM-judges reliability analysis.

\begin{figure}[H]
\centering
\includegraphics[width=0.87\linewidth]{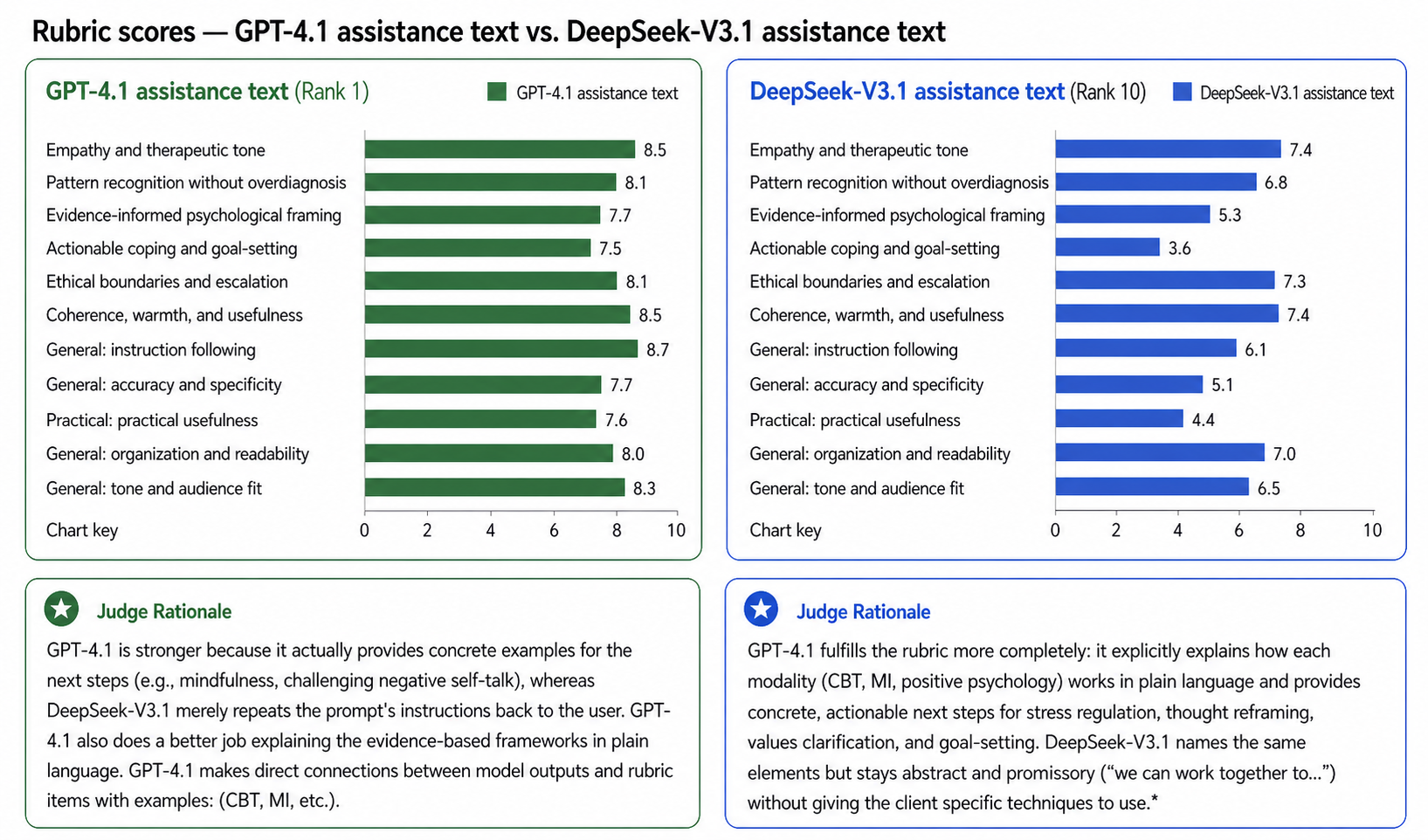}
\caption{\textbf{Rubric-guided scoring.} Judge rationales explain the pairwise choice made. The models used in this example are GPT-4.1 and DeepSeek-V3.1 in augmentation mode for the counseling task. The corresponding outputs are in Appendix~\ref{app:eval-examples}.}
\label{fig:eval-example}
\end{figure}

We aggregate the evaluations in three stages. First, each judge's choices are converted into a pairwise win rate for every model within each task and usage mode.
Second, models are ranked by win rate, and their ranks are averaged across eligible judges and ten independent runs. Third, task-level ranks are combined through a rank-of-ranks aggregation to obtain an overall ordering for each usage mode. We retain all task-level rankings, rubric scores, and rationales for subsequent analysis.

Figure~\ref{fig:eval-example} illustrates the resulting audit trail using two counseling outputs from augmentation mode. It reports the dimension-level
scores and judge rationale for outputs produced with assistance texts from GPT-4.1 and DeepSeek-V3.1. The corresponding assistance texts and final worker
outputs appear in Appendix~\ref{app:eval-examples}.

\section{Results}

We report two sets of results. First, we examine the \emph{automation} mode, where each model solves the task directly. It is imperative to acknowledge here that these results characterize models' performance on single-turn, bounded professional deliverables completed \emph{without} tool usages. This setting differs from public agentic benchmarks, which often evaluate multi-step tool use, planning, and interaction with external environments; model orderings from those benchmarks therefore need not transfer to the setting studied here.

Second, we examine the \emph{augmentation} mode, where each assistant model acts as a coach that supplies a process guidance to a fixed GPT-3.5-Turbo worker that produces the final output, rather than solving the task itself. We then compare augmentation rankings to automation rankings to test whether strong direct task-solving ability also implies strong assistant ability. Across both analyses, rankings are computed over ten independent runs of the full pipeline. Main-text figures report within-task \emph{rank of ranks} (integer ranks derived from mean ranks across runs); Appendix~\ref{app:rank-tables} reports mean rank with standard error ($\mathrm{SE}=\mathrm{SD}/\sqrt{10}$). All rankings are computed over the full candidate pool, which includes the unaided GPT-3.5-Turbo baseline. The interactive results dashboard display the full results.%
\footnote{We use the paper figures to summarize the main patterns; the \href{https://kennywong524.github.io/centaur-benchmark/}{Interactive results dashboard} provides a more detailed audit trail for readers who wish to inspect individual tasks, models, and judgments.}%

Our central result is that model performance is both task-specific and mode-specific. In automation, rankings are concentrated around a small set of strong direct solvers, although this ordering should be interpreted as specific to
the bounded, single-turn tasks evaluated here rather than as a universal ranking of model capability. In augmentation, no single assistant model dominates across all tasks: different task domains reward different assistant behaviors, and several tasks are best handled by the unaided worker baseline rather than by an external assistance. This divergence suggests that standard automation benchmarks do not capture an economically important capability, whether a model can improve another worker's output through planning and decomposition. We use the paper figures to summarize the main patterns; the interactive dashboard provides a more detailed audit trail for readers who wish to inspect individual tasks, models, and judgments.

\subsection{Model Rankings in Automation Mode}

\begin{figure}[H]
    \centering
    \includegraphics[width=\linewidth]{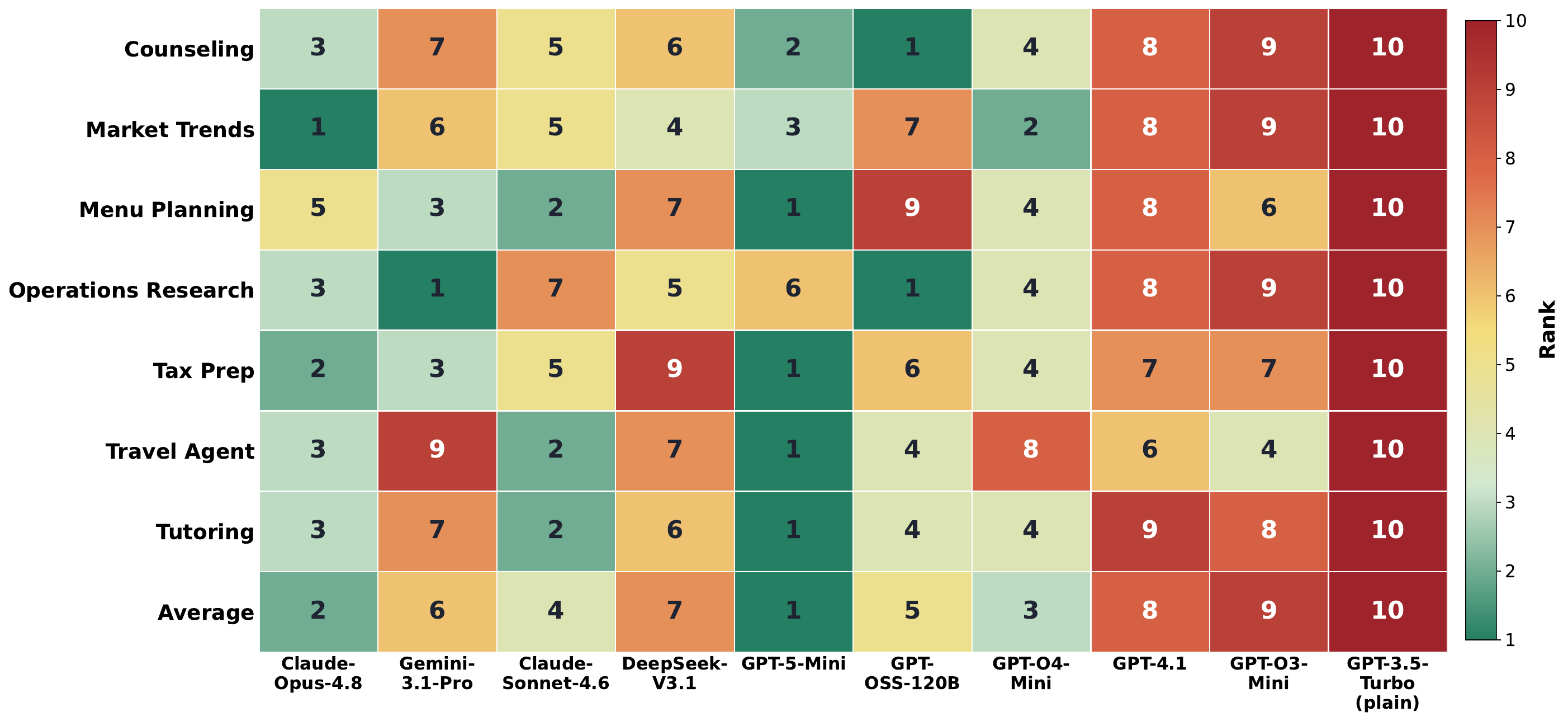}
    \caption{\textbf{Automation rankings by task, averaged across ten independent runs.} For each task, we first average each model's within-run rank across the ten runs and then rank these mean values to produce the displayed rank of ranks (1 = best). Rankings are computed using unrounded mean ranks. Exact ties receive the same rank, with the subsequent rank skipped (e.g., 1, 1, 3). Models are ordered from newest to oldest from left to right. Lower ranks indicate better performance. Mean ranks and standard errors are reported in Appendix~\ref{app:rank-tables}, Figure~\ref{fig:automation-heatmap-appendix}.}
    \label{fig:automation-heatmap}
\end{figure}

Figure~\ref{fig:automation-heatmap} reports model rankings in the automation mode by task across ten independent runs. GPT-5-Mini has the best average rank across tasks, followed by Claude-Opus-4.8. GPT-O4-Mini and Claude-Sonnet-4.6 form the next tier, while GPT-3.5-Turbo (plain) ranks last on average. The task-level results are also relatively concentrated. GPT-5-Mini ranks first in menu planning, tax preparation, travel planning, and tutoring; Claude-Opus-4.8 ranks first in market-trends analysis; GPT-OSS-120B ranks first in counseling; and GPT-OSS-120B and Gemini-3.1-Pro jointly lead operations research.

The automation results characterize direct task performance under the particular roles, tasks, and interaction constraints studied here. Automation places the full burden of task completion on the model. It must interpret the prompt, track constraints, produce the final deliverable, and avoid errors without help from another system, so general model capability matters more. We next examine whether these same models retain their relative advantages when they instead assist a fixed downstream worker.

\subsection{Model Rankings in Augmentation Mode}

\begin{figure}[H]
    \centering
    \includegraphics[width=\linewidth]{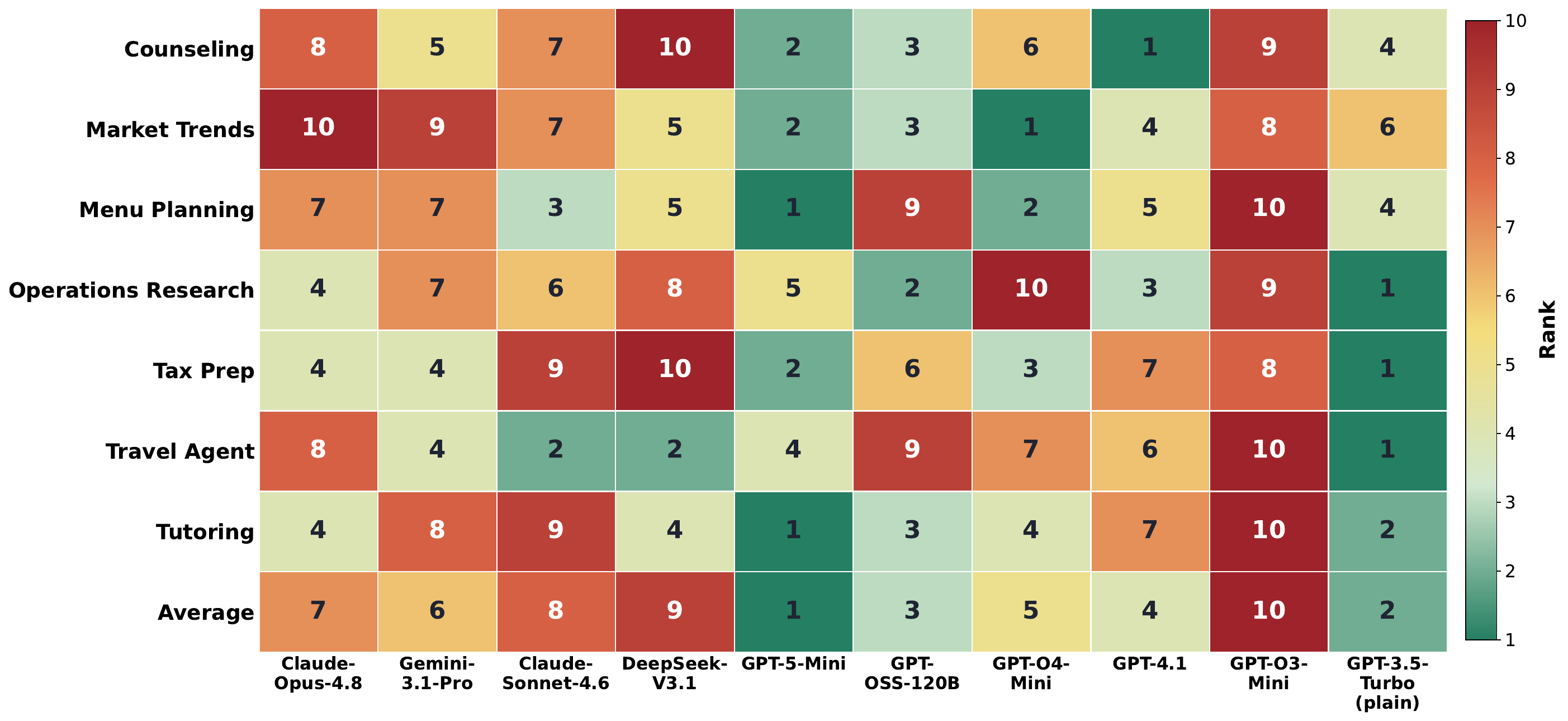}
    \caption{\textbf{Augmentation rankings by task, averaged across ten independent runs.} For each task, we first average each model's within-run rank across the ten runs and then rank these mean values to produce the displayed rank of ranks (1 = best). Rankings are computed using unrounded mean ranks. Exact ties receive the same rank, with the subsequent rank skipped (e.g., 1, 1, 3). Models are ordered from newest to oldest from left to right, with GPT-3.5-Turbo (plain), the unaided worker baseline, shown at far right. Lower ranks indicate better performance. Mean ranks and standard errors are reported in Appendix~\ref{app:rank-tables}, Figure~\ref{fig:augmentation-heatmap-appendix}.}
    \label{fig:augmentation-heatmap}
\end{figure}

Figure~\ref{fig:augmentation-heatmap} reports model rankings in the augmentation mode across ten independent runs. Each assistant model provides a process guidance in a form of an assistance text to a fixed GPT-3.5-Turbo worker, which then produces the final task output. The figure shows substantial heterogeneity across tasks: no single assistant model consistently ranks first across the task set, and the relative ordering of models shifts across counseling, tutoring, planning, and analytical tasks.

Unlike automation, the best assistant changes substantially across tasks in augmentation. GPT-4.1 ranks first in counseling, GPT-O4-Mini in market-trends analysis, GPT-5-Mini in menu planning and tutoring. However, the unaided GPT-3.5-Turbo worker baseline ranks first in operations research, tax preparation, and travel planning when it is included in the full candidate pool. The best-performing assistant therefore differs by domain rather than converging on a single model, and in some tasks assistant guidance does not improve on the fixed worker's unaided output, even worsening it.

The pattern also differs by task type. Among the assistant models, GPT-4.1 performs especially well in counseling, consistent with strength in professionally framed guidance and sensitive response structure. GPT-O4-Mini leads in market-trends analysis, where the rubric rewards concise synthesis, specificity, and actionable interpretation. GPT-5-Mini leads in menu planning and tutoring and is the best assistant in tax preparation, suggesting strength on tasks that require constraint tracking, complete deliverables, and structured explanation. GPT-OSS-120B leads among assistants in operations research, where the rubric rewards optimization framing and trade-off reasoning. Claude-Sonnet-4.6 leads among assistants in travel planning, where practical sequencing, budget awareness, and itinerary usability are central. These results suggest that different models provide different kinds of useful guidance, but the mechanisms require qualitative analysis of the guidance and the resulting worker outputs, which we return to in Section~\ref{sec:analysis}.

Importantly, these ranking differences are unlikely to be driven by models excelling on one general evaluation dimension but not another. Across the five general rubric dimensions (instruction following, accuracy and specificity, practical usefulness, organization and readability, and tone/audience fit), most models receive similar relative scores on each dimension rather than specializing on a single axis (Appendix~\ref{app:general-rubrics}, Figure~\ref{fig:general-rubric-profiles}). Augmentation rankings therefore reflect differences in overall assistant quality and task fit rather than artifacts of how the general rubric is weighted. Models do not win because they are disproportionately strong on, say, organization but weak on accuracy. The task-specific pairwise outcomes point to genuine differences in how each model guides the fixed worker.

\subsection{Automation Rankings Differ From Augmentation Rankings}
\label{sec:role-swap}

To quantify the relationship between the two usage modes, let
$r^{(k)}_{mts}$ denote the within-task rank of assistant model $m$ on task
$t$, in usage mode $s\in\{\mathrm{auto},\mathrm{aug}\}$, and independent run
$k\in\{1,\ldots,10\}$, where lower values indicate better performance. We
first average over runs,
\begin{equation}
    \bar r_{mts}=\frac{1}{10}\sum_{k=1}^{10}r^{(k)}_{mts}.
\end{equation}
For each task, we then compute
\begin{equation}
    \rho_t=\operatorname{corr}_{\mathrm{S}}
    \!\left(
      \{\bar r_{mt,\mathrm{auto}}\}_{m\in\mathcal{M}},
      \{\bar r_{mt,\mathrm{aug}}\}_{m\in\mathcal{M}}
    \right),
    \label{eq:task-mode-correlation}
\end{equation}
where $\operatorname{corr}_{\mathrm{S}}$ is Spearman's rank correlation and $\mathcal{M}$ contains the nine assistant models evaluated in both usage modes. The unaided worker is excluded because it has no corresponding assistant model in automation. For the model-level comparison, we additionally average
$\bar r_{mts}$ over the seven tasks before computing the same correlation in each usage mode. Because reverse rank is a monotonic transformation of rank, it produces the same Spearman correlations.

\begin{figure}[H]
    \centering
    \includegraphics[width=0.79\linewidth]{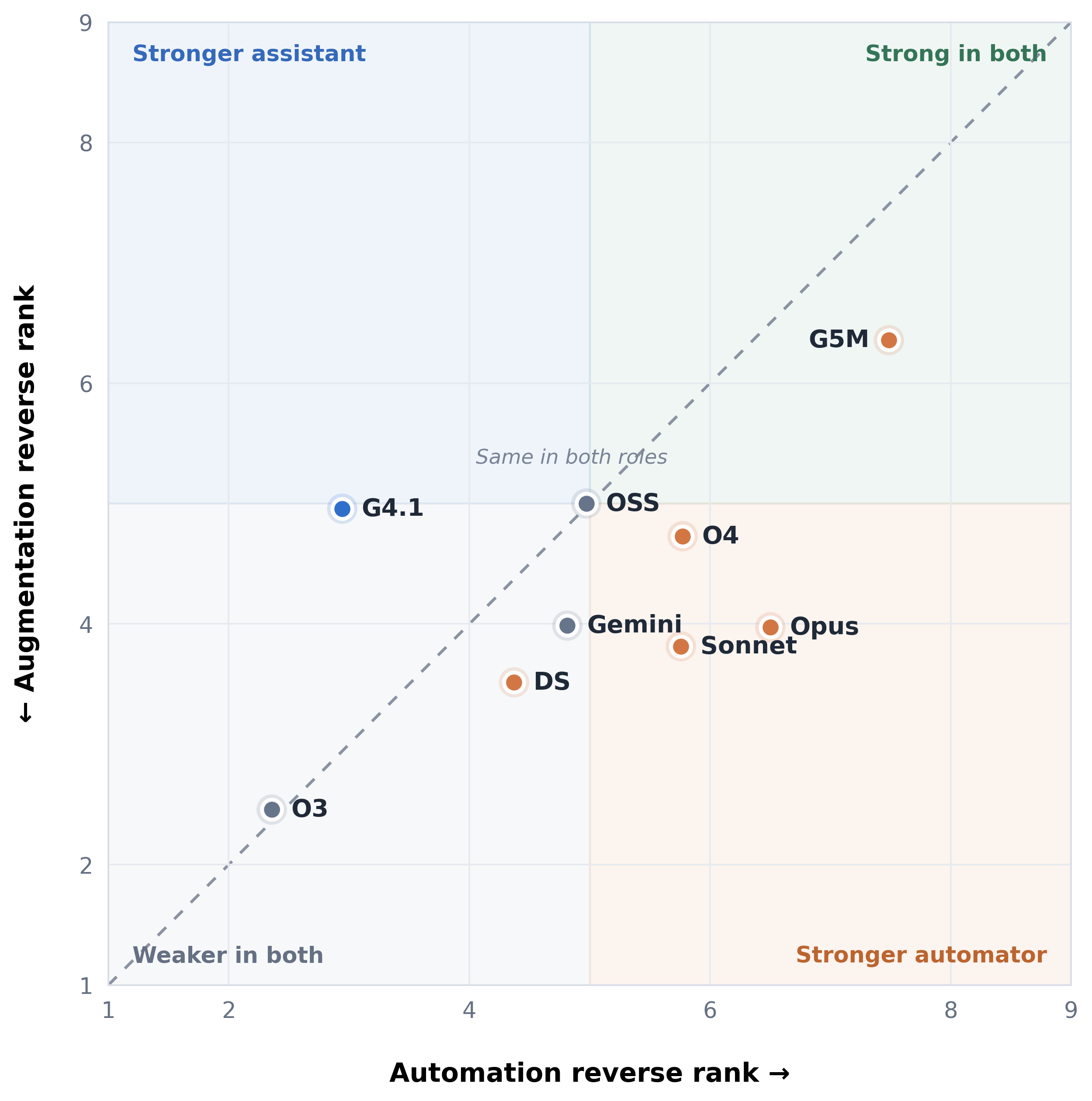}
    \caption{
    \textbf{Model-level comparison of average automation rank and average augmentation rank (assistant models only)}.
    Each point represents one assistant model's average rank across seven tasks and ten independent runs. Ranks are inverted so that higher values indicate better performance. With nine assistant models, reverse rank is defined as \(10-\bar{r}\), where \(\bar{r}\) is the model's mean rank in the corresponding usage mode. The horizontal axis reports automation performance and the vertical axis reports augmentation performance. The unaided worker baseline is excluded.}
    \label{fig:role-swap-scatter}
\end{figure}

Figure~\ref{fig:role-swap-scatter} compares each model's average automation and augmentation rank. Across the nine shared assistant models, the Spearman rank correlation between automation and augmentation performance is moderate ($\rho=0.48$; two-sided $p=0.187$), but is not statistically distinguishable
from zero at conventional significance levels. GPT-5-Mini is strong in both usage modes and has the best average rank among assistant models in augmentation as well as the best average rank in automation. GPT-4.1 is relatively stronger as an assistant than as a direct solver. Conversely, Claude-Opus-4.8, Claude-Sonnet-4.6, Gemini-3.1-Pro, and GPT-O4-Mini are stronger in automation than in augmentation. GPT-OSS-120B sits close to the diagonal. Thus, automation performance contains some information about augmentation performance, but it is an incomplete proxy for it.

\begin{table}[H]
    \centering
    \small
    \begin{tabular}{lrr}
        \toprule
        Task & Spearman $\rho_t$ & $p$-value \\
        \midrule
        Tax Preparation & 0.85 & 0.004 \\
        Menu Planning & 0.64 & 0.066 \\
        Tutoring Session Planning & 0.50 & 0.166 \\
        Counseling & 0.25 & 0.516 \\
        Operations Research Analysis & 0.17 & 0.668 \\
        Market Trend Analysis & 0.10 & 0.798 \\
        Travel Planning & -0.04 & 0.915 \\
        \bottomrule
    \end{tabular}
    \caption{\textbf{Task-level correlation between automation and augmentation rankings.} Spearman correlations are computed across the nine assistant models using each model's mean rank over ten runs. Positive values indicate similar cross-mode ordering; values near zero indicate little correspondence. $p$-values are descriptive because each task contains only nine models.}
    \label{tab:mode-correlations}
\end{table}

The strength of this correlation varies substantially by task as shown by Table~\ref{tab:mode-correlations}. Tax preparation has the strongest cross-mode correlation ($\rho_{\mathrm{tax}}=0.85$), followed by menu planning ($\rho_{\mathrm{menu}}=0.64$) and tutoring ($\rho_{\mathrm{tutor}}=0.50$). The rankings are only weakly associated in counseling ($\rho=0.25$), operations research ($\rho=0.17$), and market trends analysis ($\rho=0.10$), and are essentially unrelated in travel planning ($\rho=-0.04$). Tax preparation is the only task whose association is statistically distinguishable from zero at the conventional level ($p=0.004$), including under a Bonferroni correction for seven task-level comparisons. Broader statistical significance across tasks would require a larger model pool or additional runs, and remains an avenue for future work.

Table~\ref{tab:best-model-by-task} (Appendix~\ref{app:best-model-table}) makes the divergence concrete. It reports the best-performing model in each task under automation and augmentation, using the lowest mean rank across ten runs. In five of seven tasks, the augmentation winner differs from the automation winner. In counseling, GPT-OSS-120B is the strongest automator while GPT-4.1 is the strongest assistant; in market-trends analysis, Claude-Opus-4.8 automates best but GPT-O4-Mini assists best; and in operations research, Gemini-3.1-Pro automates best while the unaided worker baseline ranks first in augmentation. Menu planning and tutoring are the two exceptions, where GPT-5-Mini ranks first in both usage modes.

The table also shows that augmentation winners are not uniformly frontier assistant models. GPT-3.5-Turbo (plain), the unaided worker with no external assistance, ranks first in operations research, tax preparation, and travel planning. Averaged across all seven tasks, the plain baseline has a mean rank
of $3.79$; GPT-5-Mini is the only assisted condition with a better overall mean rank of $3.66$. We read this not as evidence that guidance is futile, but as
evidence that its value is task-contingent and can be negative. Poorly matched or overly complex guidance can leave the worker worse off than working alone. The corollary is central to our argument that strong direct solvers do not automatically make strong assistants, and the right model depends jointly on the role it plays and the task it supports.

Lastly, several models exhibit large and reproducible
task-level role reversals. In market trends analysis, Claude-Opus-4.8 is the strongest automator (mean rank $2.05$) but one of the weakest assistants (mean rank $8.15$). Conversely, in counseling, GPT-4.1 ranks substantially better as an assistant (mean rank $3.80$, the best among assisted conditions) than as an automator (mean rank $7.40$). These differences persist after averaging across ten independent runs and therefore are not attributable to a single random run.

The rankings establish that assistance quality is distinct from direct-solving ability, but they do not by themselves explain why particular assistance texts
help or hinder the downstream worker model. We therefore turn to the underlying assistance texts, worker outputs, rubric scores, and judge rationales.
Section~\ref{sec:analysis} examines cases with especially large gaps between ranks to identify recurring differences between effective and ineffective assistance.

\subsection{Mechanisms: Preliminary Observations on Assistance Quality}
\label{sec:analysis}

To understand what traits make for effective augmentation, we turn to the education literature to ground our observational findings. \citet{wood1976}  coined the term ``scaffold'' to define the process that enables a learner to solve a task or achieve a goal that would be beyond their unassisted efforts. In their framing, as in ours, the expert guides the learner through a task rather than completing it for them. \citet{vandepol2010} take this concept further by identifying three essential properties of effective scaffolding: (1) contingency - support is calibrated to the learner's current level, (2) fading - support is gradually withdrawn as competence grows, and (3) transfer of responsibility - the goal is learner autonomy, not dependence. The first and third properties are most directly applicable to our single-interaction setting, where scaffolds cannot adapt over time. We apply this lens to our case in order to evaluate the LLM-generated assistance texts, which are analogous to the scaffolds in education. We compare the assistance texts received by the two highest-ranked outputs against those received by the two lowest-ranked outputs across all seven tasks. In doing so, we identify three characteristics of high-quality augmentation scaffolds.

\begin{table}[H]
\centering
\definecolor{traitA}{RGB}{224,238,250}
\definecolor{traitB}{RGB}{255,247,222}
\definecolor{traitC}{RGB}{250,228,228}
\footnotesize
\setlength{\tabcolsep}{5pt}
\renewcommand{\arraystretch}{1.25}
\begin{tabularx}{\linewidth}{
    >{\raggedright\arraybackslash}p{1.0cm}
    >{\raggedright\arraybackslash}p{3.0cm}
    >{\raggedright\arraybackslash}X}
\toprule
 & \textbf{Task, assistant (rank)} & \textbf{Assistance-text excerpt} \\
\midrule

\rowcolor{traitA} \multicolumn{3}{l}{\textbf{Trait 1: Specificity beyond the prompt (contingency)}} \\
\rowcolor{traitA} \textbf{Good} & Market trends, GPT-5-Mini (1) &
``For each trend, plan 2 to 3 succinct bullets: core observation, primary driver(s), and potential market implication\ldots a short preface about methodology and data vintage.'' \\
\rowcolor{traitA} \textbf{Bad} & Market trends, GPT-4.1 (9) &
``Ensure you understand the expectation to discuss how various factors (supply, demand, weather, storage, LNG exports, infrastructure, policy) interact\ldots'' \textit{(restates the prompt's factor list)} \\
\addlinespace

\rowcolor{traitB} \multicolumn{3}{l}{\textbf{Trait 2: Explicit structure and ordering}} \\
\rowcolor{traitB} \textbf{Good} & Travel planning, Gemini-3.1-Pro (3) &
Enforces assumptions and clarifying questions ``\emph{before} the provisional itinerary,'' with named sections in explicit order: Introduction $\to$ Budget Breakdown $\to$ Daily Itinerary $\to$ Conclusion. \\
\rowcolor{traitB} \textbf{Bad} & Travel planning, GPT-O4-Mini (9) &
``Determine sequence and hierarchy of information for clarity and flow''; instructs worker to ``populate each section with neutral placeholders\ldots\ \emph{[Day 1: transportation placeholder]}''; no ordering is prescribed. \\
\addlinespace

\rowcolor{traitC} \multicolumn{3}{l}{\textbf{Trait 3: Not working against the task}} \\
\rowcolor{traitC} \textbf{Good} & Operations research, Gemini-3.1-Pro (3) &
``Present two distinct, practical strategies. For each, explicitly state the expected benefits, potential drawbacks, and associated risks.'' \\
\rowcolor{traitC} \textbf{Bad} & Operations research, GPT-O3-Mini (9) &
``Establish a workflow for integrating\ldots trade-offs \textbf{without delving into specific solutions or analysis}''; finalize ``\textbf{without\ldots task-specific analytical content}.'' \\
\addlinespace

\bottomrule
\end{tabularx}
\caption{\textbf{Illustrative higher- vs.\ lower-ranked assistance texts for each
scaffold-quality characteristic.} Rows are color-coded by trait; ranks are the aggregate
augmentation rank within the task run. Excerpts are drawn from the assistance texts and lightly trimmed.}
\label{tab:trait-examples}
\end{table}

First, good guidance includes specificity that adds analytical value beyond the prompt, rather than just regurgitating information from the prompt, which is an application of the contingency principle, where effective scaffolding is calibrated to the learner's specific challenge rather than offered generically \citep{vandepol2010}. Table~\ref{tab:trait-examples} illustrates each trait with a higher- and lower-ranked assistance text. Low-performing guidance messages that simply echo the prompt's own constraints do not augment the focal worker, who already have access to the prompt. For example, in the market trends task, a top assistance text directs the worker to pair each trend with its primary driver and likely market implication and to note the underlying methodology, whereas a bottom-tier assistance text merely restates the prompt's own list of factors without adding an analytical step.

Next, embedding explicit structural requirements into the workflow is another trait of a top assistance text. In the travel planning task, a top assistance text enforces assumptions and clarifying questions before the provisional itinerary and prescribes a named section sequence (introduction, budget breakdown, day-by-day itinerary, and conclusion), making the ordering non-negotiable. A bottom-tier assistance text, by contrast, delegates ordering to the worker by instructing it to ``determine sequence and hierarchy of information for clarity and flow,'' and substitutes neutral placeholders for actual structural guidance.

Third, a negative point that good scaffolds avoid: actively working against task completion. Effective scaffolding includes what \citet{wood1976} term \textit{direction maintenance}, keeping the learner oriented toward the task objective. The most harmful assistance texts in our study invert this function entirely: rather than correcting learner drift, the scaffold itself redirects the worker away from the core deliverable, by instructing the worker not to finish the task as described in the prompt. In operations research task, a bottom ranked scaffold instructs the worker to establish a workflow for integrating necessary performance indicators, risk discussions, and balanced trade-offs without delving into specific solutions or analysis, yet the task explicitly requires two practical solutions with trade-offs stated. This instruction directly prohibits the core deliverable. Thus, no assistance can often beat bad assistance. Notably, the unaided worker is a strong baseline: the plain condition (GPT-3.5-Turbo with no assistance) ranks at or near the top in the many tasks. Poorly calibrated assistance is therefore frequently worse than none, undermining the transfer of responsibility that effective scaffolding aims to achieve \citep{vandepol2010}.

\section{Discussion}

Our results show that automation and augmentation are not two labels for the same underlying capability. They produce systematically different rankings. The top-ranked model differs between automation and assistant-only augmentation in five of the seven tasks. Thus, the assistant value is sharply task-contingent. Interpreting these patterns requires moving beyond leaderboard thinking to role-specific model selection. The automation results are comparatively concentrated around a small set of
direct solvers, while GPT-3.5-Turbo ranks last. This ordering characterizes performance on our single-turn professional deliverables and should not be interpreted as reproducing the orderings of agentic benchmarks built
around multi-step tool use and environment interaction.
Augmentation tells a different story. No single model dominates across tasks, average ranks are more dispersed, and several strong direct solvers perform poorly as assistants. Most strikingly, on operations research, tax preparation, and travel planning, the unaided GPT-3.5-Turbo worker outranks every assisted condition, and it has the second best average augmentation rank overall after GPT-5-Mini. We read this not as evidence that assistance is useless, but as evidence that its value is conditional and can be negative. Poorly matched or overly complex process guidance can distract the worker, constrain useful execution, or push it away from task requirements.

\subsection{Implications for Organizations and Workflow Design}

Our framework offers several practical applications for individuals and organizations adopting AI systems. First, it can support model selection across teams, products, and workflows. Teams seeking a reliable solver for well-defined, repeatable tasks may prioritize automation performance, whereas teams embedding AI into expert workflows, where the goal is to improve human output rather than replace human judgment, may prioritize augmentation ability.

Importantly, the framework is modular. In our experiments, we use GPT-3.5-Turbo as an illustrative fixed worker model to demonstrate the pipeline and hold the worker role constant across augmentation trials. However, individuals and organizations can substitute this worker model with a model, human baseline, or internal workflow proxy that better reflects their own deployment context. For example, a consulting firm could evaluate how different assistant models guide junior analysts, domain experts, customer-support agents, or a preferred internal model. This flexibility allows the framework to be adapted to local goals, constraints, and performance standards.

Secondly, designing human-AI collaboration can be done more efficiently and effectively. Beyond model selection, our framework can guide how organizations structure the division of labor between humans and AI. At the organizational level, different models can be assigned to different subtasks according to where each model's comparative advantage lies. At the team level, human-AI pairs can use the automation/augmentation profiles to decide which subtasks to delegate entirely versus which to treat as collaborative, keeping humans in the loop where an assistance yields the largest gains. At the individual level, a practitioner can use the same logic to choose the tool best matched to their specific workflow, rather than defaulting to a single general-purpose assistant.

\subsection{Implications for Evaluation and Research}

Our findings also speak to how LLMs should be evaluated. Many existing benchmarks measure automation almost exclusively. If gains on those benchmarks do not transfer to augmentation, then labs and leaderboard builders may be optimizing for only one economically relevant capability. Adopting automation and augmentation as distinct evaluation axes would make model profiles more informative for human-in-the-loop and multi-agent deployment. If automation and augmentation are empirically distinct capabilities, then any evaluation that tests only one is incomplete by design, and our framework offers a template for fixing that. Leaderboard builders and AI labs optimizing for benchmark performance could adopt the automation vs. augmentation distinction as a structural requirement for capability evaluation.

CentaurBench sits between two literatures. Relative to large CS benchmarks, it evaluates economically grounded occupational tasks under a fixed coaching protocol rather than abstract capability alone. Relative to field experiments, it compares many frontier models across both roles and multiple domains rather than a single tool in a single setting. Relative to marketplace benchmarks such as HAPI \citep{upwork_hapi2025}, it holds the downstream worker fixed to isolate assistant quality across models at scale. The appropriate next step is not to replace one paradigm with another, but to combine them for optimal model selection.

\subsection{Limitations and Future Work}

This study introduces a simulation and evaluation framework to measure models' ability to augment and automate tasks across usage modes. We present this as an initial pilot application; it should not be interpreted as a definitive mapping
of model capabilities across tasks and usage modes. The results identify
settings in which automation and augmentation rankings diverge and provide
preliminary model profiles for subsequent validation. More broadly, the study
motivates the development of assistance-oriented benchmarks rather than
claiming to provide a comprehensive benchmark itself.

Because the current evaluation relies on LLM judges, the findings constitute
scalable simulation evidence rather than definitive human or domain-expert
judgments. Although we use general and task-specific rubrics, multiple judges,
leave-family-out evaluation, and repeated runs, these procedures cannot eliminate all evaluator bias. Human validation remains a priority, particularly for tasks such as counseling, tax preparation,
tutoring, and operations research, where domain expertise and professional
judgment are consequential to assess whether assistant advantages observed in simulated settings persist in authentic workplace contexts.

Several extensions are especially important. First, the present design evaluates augmentation through a single assistance text passed to a fixed worker. Real professional workflows are often iterative: assistants clarify requirements,
revise plans, and respond to feedback across multiple turns. Extending the
framework to multi-turn and agentic settings, in which an assistant contributes
throughout an evolving workflow rather than through one-shot guidance, is a
natural next step.

Second, all augmentation conditions use the same worker model, GPT-3.5-Turbo. Holding the worker fixed isolates variation in the assistance
provided, but it also limits generalizability. Guidance that helps one worker may not transfer to another. A stronger or differently aligned worker may need less assistance, while a weaker worker may benefit more from structured
guidance. Future studies should vary worker capability and model family to test
whether assistant rankings remain stable across downstream worker models.

Third, the pilot includes seven professional tasks and ten independent
generation-and-evaluation runs. Replication improves the stability of the
reported estimates, but the task set remains modest relative to the diversity
of real-world, occupational tasks. Expanding the evaluation to additional tasks, industries, risk levels, and deliverable formats would provide stronger evidence about the generalizability of role-specific model performance.

External validation should combine domain expertise with authentic workplace
tasks. Upwork's Human+Agent Productivity Index (HAPI) \citep{upwork_hapi2025}, for example, emphasizes verified marketplace
deliverables and human rubric scoring. Our design instead holds the downstream
worker fixed to isolate differences in assistance quality across models at
scale. Future work could combine these strengths by using workplace-grounded
tasks, validating a preregistered subset with domain experts, and explicitly
comparing automation and augmentation rankings. This would establish whether
the preferences of LLM judges correspond to expert assessments of final-output quality, guidance usefulness, and workplace value.

The best mathematicians are not necessarily the best mathematics teachers, and the highest-ranked tennis players are not necessarily the best coaches. As AI
systems increasingly advise humans and coordinate with other models, evaluation
must measure not only a model's ability to automate tasks, but also its capacity
to augment the performance of others. This study offers an initial framework
for making that distinction measurable and motivates a broader evaluation
agenda organized around the role a model plays, the worker it supports, and the
task they jointly perform.

More broadly, realizing the potential of AI to complement human judgment, expand human capability, and support human
flourishing will require measures of progress that extend beyond autonomous
task completion. Our framework represents one step towards reframing AI
progress around not only what systems can accomplish independently, but also
how effectively they enable humans and other agents to accomplish even more.

\bibliographystyle{plainnat}
\bibliography{references}

\appendix
\section{Models and Evaluators}
\label{app:models}
Table~\ref{tab:models} lists the assistant models evaluated in automation and augmentation, the fixed worker model used in augmentation, and the LLM-as-judge panel. Model families are grouped to match the leave-one-family-out judging rule described in Section~\ref{sec:methodology}.

\begin{table}[htbp]
\centering
\renewcommand{\arraystretch}{1.2}
\begin{tabular}{llccc}
\toprule
\textbf{Model} & \textbf{Provider} & \textbf{Automation} & \textbf{Augmentation} & \textbf{Judge} \\
\midrule
GPT-3.5-Turbo     & OpenAI    & \checkmark & \textit{Worker} & \\
\midrule
Claude-Opus-4.8   & Anthropic & \checkmark & \checkmark      & \checkmark \\
Claude-Sonnet-4.6 & Anthropic & \checkmark & \checkmark      & \\
\midrule
DeepSeek-V3.1     & DeepSeek  & \checkmark & \checkmark      & \checkmark \\
\midrule
Gemini-3.1-Pro    & Google    & \checkmark & \checkmark      & \checkmark \\
\midrule
GPT-4.1           & OpenAI    & \checkmark & \checkmark      & \checkmark \\
GPT-OSS-120B      & OpenAI    & \checkmark & \checkmark      & \\
GPT-5-Mini        & OpenAI    & \checkmark & \checkmark      & \\
GPT-O4-Mini       & OpenAI    & \checkmark & \checkmark      & \\
GPT-O3-Mini       & OpenAI    & \checkmark & \checkmark      & \\
\bottomrule
\end{tabular}
\caption{\textbf{Models used in the evaluation and their assigned roles.} All focal models serve as direct solvers in the automation regime and as planning assistants in the augmentation regime. GPT-3.5-Turbo serves as the fixed worker in augmentation mode rather than as a focal model. Judges follow a leave-family-out constraint: each judge is excluded from evaluating outputs of models from its own provider family.}
\label{tab:models}
\end{table}

\section{Task Prompts and Evaluation Rubrics}
\label{app:prompts}

\tcbset{
  systemprompt/.style={
    colback=gray!6!white,
    colframe=gray!45!black,
    colbacktitle=gray!45!black,
    coltitle=white,
    fonttitle=\bfseries\small,
    boxrule=0.5pt,
    arc=3pt,
    breakable,
    left=6pt, right=6pt, top=4pt, bottom=4pt,
    toptitle=3pt, bottomtitle=3pt
  }
}

\tcbset{
  taskprompt/.style={
    colback=blue!4!white,
    colframe=blue!35!black,
    colbacktitle=blue!35!black,
    coltitle=white,
    fonttitle=\bfseries\small,
    boxrule=0.5pt,
    arc=3pt,
    breakable,
    left=6pt, right=6pt, top=4pt, bottom=4pt,
    toptitle=3pt, bottomtitle=3pt
  }
}

\tcbset{
  rubricbox/.style={
    colback=teal!4!white,
    colframe=teal!40!black,
    colbacktitle=teal!40!black,
    coltitle=white,
    fonttitle=\bfseries\small,
    boxrule=0.5pt,
    arc=3pt,
    breakable,
    left=6pt, right=6pt, top=4pt, bottom=4pt,
    toptitle=3pt, bottomtitle=3pt
  }
}

All task prompts and rubrics are reproduced in full below.
Each task is evaluated using the universal judge prompt together
with the task-specific rubric dimensions listed here.
The augmentation condition additionally uses the universal
assistance text prompt to generate planning guidance for the worker model.

\subsection{Universal Assistance Text Prompt}
\label{appendix:scaffold}

\begin{tcolorbox}[systemprompt, title=Worker Instruction]
You receive internal assistant guidance plus a client task.
Use the guidance for planning only; never copy or restate it.
Complete the client task fully and return only the complete final deliverable.
\end{tcolorbox}

\vspace{0.5em}

\begin{tcolorbox}[systemprompt, title=Assistance Text Prompt (Assistant $\to$ Worker)]
You are a planning coach helping another model prepare to complete a task.

Your role is to provide \textbf{process guidance only}, not task content.

Write a short 200--250 word guidance message that gives a three-phase workflow:

\begin{enumerate}
  \setlength{\itemsep}{2pt}
  \item \textbf{Requirements Check}
    \begin{itemize}
      \setlength{\itemsep}{0pt}
      \item Identify the deliverable, audience, tone, and constraints.
      \item Note what a strong answer must include.
    \end{itemize}
  \item \textbf{Plan \& Structure}
    \begin{itemize}
      \setlength{\itemsep}{0pt}
      \item Suggest a general response structure or sequence of sections.
      \item Suggest how to balance completeness, clarity, and tone.
    \end{itemize}
  \item \textbf{Draft $\to$ Self-Review $\to$ Finalize}
    \begin{itemize}
      \setlength{\itemsep}{0pt}
      \item Suggest a brief checklist for revising the answer before submission.
    \end{itemize}
\end{enumerate}

\textbf{Important rules:}
\begin{itemize}
  \setlength{\itemsep}{0pt}
  \item Do \textbf{not} solve the task.
  \item Do \textbf{not} provide topic-specific content, examples, or recommendations.
  \item Do \textbf{not} write sentences that could be copied into the final answer.
  \item Do \textbf{not} restate the task as a completed outline with filled-in content.
  \item Keep the guidance generic and process-focused.
\end{itemize}

Format in Markdown under the heading:
\texttt{**Assistant Guidance: Three-Phase Workflow**}
\end{tcolorbox}

\subsection{Universal Judge Prompt}
\label{appendix:judge}

\begin{tcolorbox}[systemprompt, title=Universal Judge Prompt]
You are an expert evaluator comparing two candidate responses to the same task.

First, silently score each option from 1--10 on each rubric dimension below.
A score of 1 means the response fails that dimension; 10 means it satisfies it
excellently. Use the task-specific dimensions as primary criteria and the general
dimensions as tie-breakers.

\textbf{Primary rule:} prefer the response that would be more useful, accurate,
complete, and reliable for the intended user. Do not reward verbosity unless it
improves usefulness.

\textbf{Task-specific rubric:} {[insert task-specific rubric below]}

\textbf{General rubric:}
\begin{enumerate}
  \setlength{\itemsep}{0pt}
  \item Instruction following: satisfies all explicit requirements and constraints.
  \item Accuracy and specificity: avoids false claims, vague filler, and unsupported
        assumptions.
  \item Practical usefulness: gives concrete, usable guidance or outputs.
  \item Organization and readability: clear structure, easy to scan, appropriate
        formatting.
  \item Tone and audience fit: matches the role, user need, and professional context.
\end{enumerate}

After scoring, choose the better option overall based on the average score across all
rubrics for both options. If both are flawed, choose the one with fewer serious
errors. If one response violates a hard safety, legal, medical, or dietary constraint,
strongly prefer the other response unless it has an equally serious issue.

\textbf{Return only:} the final choice (\texttt{Option\_1} or \texttt{Option\_2}),
the score for each rubric dimension (1--10) for both options, and the average score
across all rubrics for both options. Structure the output clearly with headings for
the final choice, all rubric scores, and the average score.
\end{tcolorbox}

\subsection{Counseling}
\label{appendix:counseling}

\begin{tcolorbox}[taskprompt, title=Task Prompt — Counseling]
Write a single-session supportive counseling-style response for a client who feels
emotionally drained, anxious about work, and uncertain about life direction.
The response must include:
\begin{enumerate}
  \setlength{\itemsep}{0pt}
  \item Empathic validation of the client's feelings without minimizing them.
  \item A cautious identification of possible patterns such as burnout,
        perfectionism, avoidance, low self-efficacy, or value-direction mismatch,
        without diagnosing.
  \item Evidence-informed framing using CBT, motivational interviewing, and/or
        positive psychology in plain language.
  \item Concrete next steps for stress regulation, thought reframing, values
        clarification, and near-term goal setting.
  \item A brief safety/limits note encouraging professional support if distress is
        severe or persistent.
  \item An encouraging close that reinforces agency and hope.
\end{enumerate}
\end{tcolorbox}

\begin{tcolorbox}[rubricbox, title=Evaluation Rubric — Counseling]
Score each dimension from 1 (lowest) to 10 (highest):
\begin{enumerate}
  \setlength{\itemsep}{0pt}
  \item Empathy and therapeutic tone.
  \item Appropriate pattern recognition without overdiagnosis.
  \item Evidence-informed psychological framing.
  \item Actionable coping and goal-setting recommendations.
  \item Ethical boundaries and escalation guidance.
  \item Coherence, warmth, and usefulness as a single-session response.
\end{enumerate}
\end{tcolorbox}

\subsection{Travel Planning}
\label{appendix:travel}

\begin{tcolorbox}[taskprompt, title=Task Prompt — Travel Planning]
Create a 5-day Tokyo itinerary for one traveler next month with a total budget of
\$2,500 USD, excluding passport costs. The traveler is departing from San Francisco.
Clearly state assumptions and give clarifying questions before the provisional plan.
Use realistic estimates, label them as estimates, and keep the plan budget-aware.
\end{tcolorbox}

\begin{tcolorbox}[rubricbox, title=Evaluation Rubric — Travel Planning]
Score each dimension from 1 (lowest) to 10 (highest):
\begin{enumerate}
  \setlength{\itemsep}{0pt}
  \item Completeness: covers questions, costs, hotels, transport,
        customs/regulations, itinerary, and budget confirmation.
  \item Cost realism and arithmetic: flights, lodging, transport, food/activities,
        and total are plausible and correctly summed. Penalize outputs that exceed
        the budget.
  \item Hotel and transport practicality: viable hotel options, airport/transit
        guidance, rail/pass advice.
  \item Itinerary quality: 5 days with morning/afternoon/evening plans that are
        geographically sensible and well-optimized for distance and fatigue.
  \item Travel-agent professionalism: clear, customer-focused, with appropriate
        caveats around estimates.
  \item Handling uncertainty: asks clarifying questions for missing information
        while still providing a useful provisional plan.
\end{enumerate}
\end{tcolorbox}

\subsection{Meal Planning}
\label{appendix:meal}

\begin{tcolorbox}[taskprompt, title=Task Prompt — Meal Planning]
Create a cheap, simple 7-day meal plan for a 31-year-old Australian adult with
shellfish and gluten allergies, lactose and onion intolerance, and preferences for
meat, potatoes, Mexican, Thai, and Indian flavors. Avoid very salty or very sweet
foods and account for super-taster sensitivity by using mild, adjustable seasoning.
Include breakfast, lunch, dinner, and simple snacks for each day, plus a grocery
list and brief prep notes. Do not include shellfish, gluten-containing ingredients,
lactose-containing dairy, or onion.
\end{tcolorbox}

\begin{tcolorbox}[rubricbox, title=Evaluation Rubric — Meal Planning]
Score each dimension from 1 (lowest) to 10 (highest):
\begin{enumerate}
  \setlength{\itemsep}{0pt}
  \item Dietary safety: strictly avoids shellfish, gluten, lactose, and onion.
  \item Preference fit: uses liked cuisines and foods while avoiding very
        salty/sweet flavors.
  \item Affordability: grocery list items are readily available at common grocery
        stores and not excessively expensive.
  \item Nutritional adequacy and variety: the plan is not overly repetitive and
        covers proteins, carbohydrates, vegetables, and fruits.
  \item Simplicity and usability: easy to follow and clearly structured for the
        target audience.
  \item Completeness: covers breakfast, lunch, dinner, and snacks for each day,
        plus a grocery list and prep notes. Penalize any missing components.
\end{enumerate}
\end{tcolorbox}

\subsection{Tax Preparation}
\label{appendix:tax}

\begin{tcolorbox}[taskprompt, title=Task Prompt — Tax Preparation]
Assume tax year 2025 and state: California. A client brings partially completed federal and California returns. Review the provided data, identify discrepancies, explain the applicable federal/state rules, recalculate key figures using stated assumptions, and explain what forms/schedules need correction. If information is insufficient for an exact tax liability, state what is missing and provide a reasonable estimate or calculation framework.

\medskip
\textbf{Client-provided facts:}
\begin{itemize}
  \setlength{\itemsep}{1pt}
  \item W-2 wages: \$68,500.
  \item Freelance/contract income: \$4,200 reported on Form 1099-NEC.
  \item Business expenses for freelance work: not yet provided.
  \item Mortgage interest paid on Form 1098: \$9,800.
  \item California property tax paid: \$4,800.
  \item California income tax withheld: \$2,900.
  \item One dependent listed: a 22-year-old full-time college student who lived with the client for 9 months. The student earned \$18,000 from a part-time job; the client says the student used most earnings for savings and personal expenses, and the client paid more than half of household costs and more than half of the student's total support.
  \item Filing status selected by the client: Single.
\end{itemize}

\medskip
\textbf{Pre-filled portions of the client's federal and California returns:}
\begin{itemize}
  \setlength{\itemsep}{1pt}
  \item Form 1040 includes W-2 wages of \$68,500.
  \item Form 1040 includes freelance income of \$2,400 instead of the \$4,200 shown on the 1099-NEC.
  \item No Schedule~C has been prepared.
  \item No Schedule~SE has been prepared, and no self-employment tax appears on Schedule~2.
  \item Form 1040 Line~12 shows a deduction of \$9,800 labeled as the standard deduction.
  \item No Schedule~A has been prepared even though mortgage interest, property tax, and state tax withholding are available.
  \item The dependent is claimed as qualifying for the Child Tax Credit.
  \item The return does not analyze whether Head of Household filing status is available.
  \item California Form 540 shows federal AGI carried over as \$62,000.
  \item California tax was computed using the uncorrected federal income figures.
\end{itemize}

\medskip
All calculations were done by the client before coming to you. Do not assume the pre-filled entries are correct.
\end{tcolorbox}

\begin{tcolorbox}[rubricbox, title=Evaluation Rubric — Tax Preparation]
Score each dimension from 1 (lowest) to 10 (highest):
\begin{enumerate}
  \setlength{\itemsep}{0pt}
  \item Rule accuracy: correctly applies federal and California rules regarding
        self-employment income, dependent rules, mortgage interest, standard
        deduction, and state/federal differences without seeking loopholes.
  \item Error detection: there are 10 expected discrepancies/issues. Score according to how many are detected (10 detected $= 10$; 2--9 detected $=$ that number; 0--1 detected $= 1$). Expected issues:
      \begin{enumerate}
        \setlength{\itemsep}{0pt}
        \item[\textit{a.}] Catches the \$2,400 vs.\ \$4,200 freelance income mismatch.
        \item[\textit{b.}] Identifies missing Schedule~C.
        \item[\textit{c.}] Identifies missing Schedule~SE\slash self-employment tax.
        \item[\textit{d.}] Identifies that \$9,800 mortgage interest is not the standard deduction.
        \item[\textit{e.}] Identifies missing Schedule~A\slash itemization comparison.
        \item[\textit{f.}] Identifies improper Child Tax Credit treatment for a 22-year-old.
        \item[\textit{g.}] Analyzes possible Credit for Other Dependents instead.
        \item[\textit{h.}] Analyzes whether Single should be changed to Head of Household given the dependent and household-cost facts.
        \item[\textit{i.}] Catches California AGI carryover inconsistency.
        \item[\textit{j.}] Catches California tax computed from uncorrected federal figures.
      \end{enumerate}
  \item Calculation quality: provides income, AGI, deductions, taxable income,
        self-employment tax, and estimates where possible.
  \item Form guidance: clear guidance on Schedule~C, Schedule~SE, Schedule~A vs.\
        standard deduction, and Form~1040 placement.
  \item Dependent analysis: handles the 22-year-old full-time student with
        \$18{,}000 income cautiously and correctly.
  \item Clear communication: translates dense tax rules into plain English so the
        client understands what they owe and why.
\end{enumerate}
\end{tcolorbox}

\subsection{Tutoring}
\label{appendix:tutoring}

\begin{tcolorbox}[taskprompt, title=Task Prompt — Tutoring]
Prepare a classroom-ready Grade~3 lesson segment for 8-year-olds on improper
fractions and mixed numbers. Include a clear learning objective, simple explanation,
analogy, worked examples, common misconceptions, interactive activity, checks for
understanding, and a short wrap-up. Use age-appropriate language and avoid abstract
notation without visual support.
\end{tcolorbox}

\begin{tcolorbox}[rubricbox, title=Evaluation Rubric — Tutoring]
Score each dimension from 1 (lowest) to 10 (highest):
\begin{enumerate}
  \setlength{\itemsep}{0pt}
  \item Mathematical correctness: accurate representation of concepts with no
        errors. Penalize any incorrect examples or inaccurate concepts.
  \item Age-appropriate explanation and pacing: clear, concise, logically ordered,
        and avoids abstract notation unsuitable for 8-year-olds.
  \item Analogy quality: appropriate and accessible analogy, not too abstract for
        the target age group.
  \item Misconception handling: clearly identifies common errors, debunks them
        effectively, and provides easy-to-remember strategies for students.
  \item Classroom practicality and engagement: designed for immediate classroom
        use with interactive elements that maintain student interest.
  \item Assessment and checks for understanding: includes formative assessment
        techniques such as thumbs up/down, exit tickets with a visual fraction
        problem, or think-pair-share.
\end{enumerate}
\end{tcolorbox}

\subsection{Operations Research}
\label{appendix:or}

\begin{tcolorbox}[taskprompt, title=Task Prompt — Operations Research]
You are an Operations Research Analyst at a national logistics firm that manages
inventory and delivery for hundreds of retail stores. Delivery delays and rising
fuel costs are worrying senior management. They have asked you to identify the key
operational bottlenecks and propose data-driven strategies to improve overall
efficiency.

Your deliverable is a concise internal report (300--400 words) for executives that:
\begin{itemize}
  \setlength{\itemsep}{0pt}
  \item Defines the data to collect and how you would validate it.
  \item Outlines an analytical/optimization framework to evaluate alternate
        delivery schedules or warehouse allocations.
  \item Recommends at least two practical solutions that balance cost, service
        levels, and resource constraints, with trade-offs and risks stated.
  \item Specifies KPIs to monitor and explains how results would be communicated
        to management.
\end{itemize}
Write clearly, with headings and bullet points where useful.
\end{tcolorbox}

\begin{tcolorbox}[rubricbox, title=Evaluation Rubric — Operations Research]
Score each dimension from 1 (lowest) to 10 (highest):
\begin{enumerate}
  \setlength{\itemsep}{0pt}
  \item Data and validation: identifies route, demand, fleet, warehouse, service,
        and cost data; explains collection techniques and validation methods.
  \item Analytical framework: proposes suitable OR methods such as MILP,
        simulation, scenario analysis, or routing optimization, formulated clearly
        enough to be tractable.
  \item Recommendations: at least two practical solutions addressing bottlenecks,
        with clear lines of action.
  \item Trade-offs and risks: acknowledges pitfalls and states cost, service, and
        resource implications explicitly without overconfidence.
  \item KPIs and communication: clear metrics and an executive reporting plan
        accessible to both technical and non-technical stakeholders.
  \item Executive memo quality: concise, within the 300--400 word limit,
        well-structured, self-contained, and professionally toned.
\end{enumerate}
\end{tcolorbox}

\subsection{Market Trends Analysis}
\label{appendix:market}

\begin{tcolorbox}[taskprompt, title=Task Prompt — Market Trends Analysis]
Imagine you are briefing a client on the U.S.\ energy market in 2026. Identify
exactly three positive or bullish trends and exactly three negative or bearish trends
likely to influence natural gas prices this year. Discuss briefly how factors like
supply, demand, weather, storage, LNG exports, infrastructure, and policy interact
to shape these trends. Conclude with 2--3 sentences summarizing the overall outlook
for investors or companies in the energy sector.
\end{tcolorbox}

\begin{tcolorbox}[rubricbox, title=Evaluation Rubric — Market Trends Analysis]
Score each dimension from 1 (lowest) to 10 (highest):
\begin{enumerate}
  \setlength{\itemsep}{0pt}
  \item Trend completeness and balance: identifies exactly three bullish and
        exactly three bearish trends. Penalize missing, extra, or poorly
        categorized trends.
  \item Economic and market accuracy: correctly explains natural gas market
        drivers including supply, demand, storage, LNG exports, weather,
        infrastructure, and policy. Penalize fabricated statistics or
        implausible causal logic.
  \item Causal reasoning and interaction effects: explains how drivers interact
        rather than listing trends in isolation.
  \item Client usefulness and actionability: provides insights useful to investors
        or energy-sector firms, including practical implications and risks.
  \item Specificity without overclaiming: gives grounded analysis while
        appropriately caveating uncertainty. Penalize vague filler or confident
        forecasts without basis.
  \item Conclusion quality: ends with a clear 2--3 sentence outlook that
        synthesizes bullish and bearish pressures for the intended audience.
\end{enumerate}
\end{tcolorbox}

\section{Rank Tables With Standard Error}
\label{app:rank-tables}

\paragraph{Construction of uncertainty estimates.}
For each task--model cell, we report the mean rank across the ten independent
runs and its standard error across those ten observations. Specifically, for each model we first average its ranks across the seven tasks within each run, yielding ten run-level average ranks, and then calculate the standard error across those ten values.

\begin{figure}[H]
    \centering
    \includegraphics[width=1.02\linewidth]{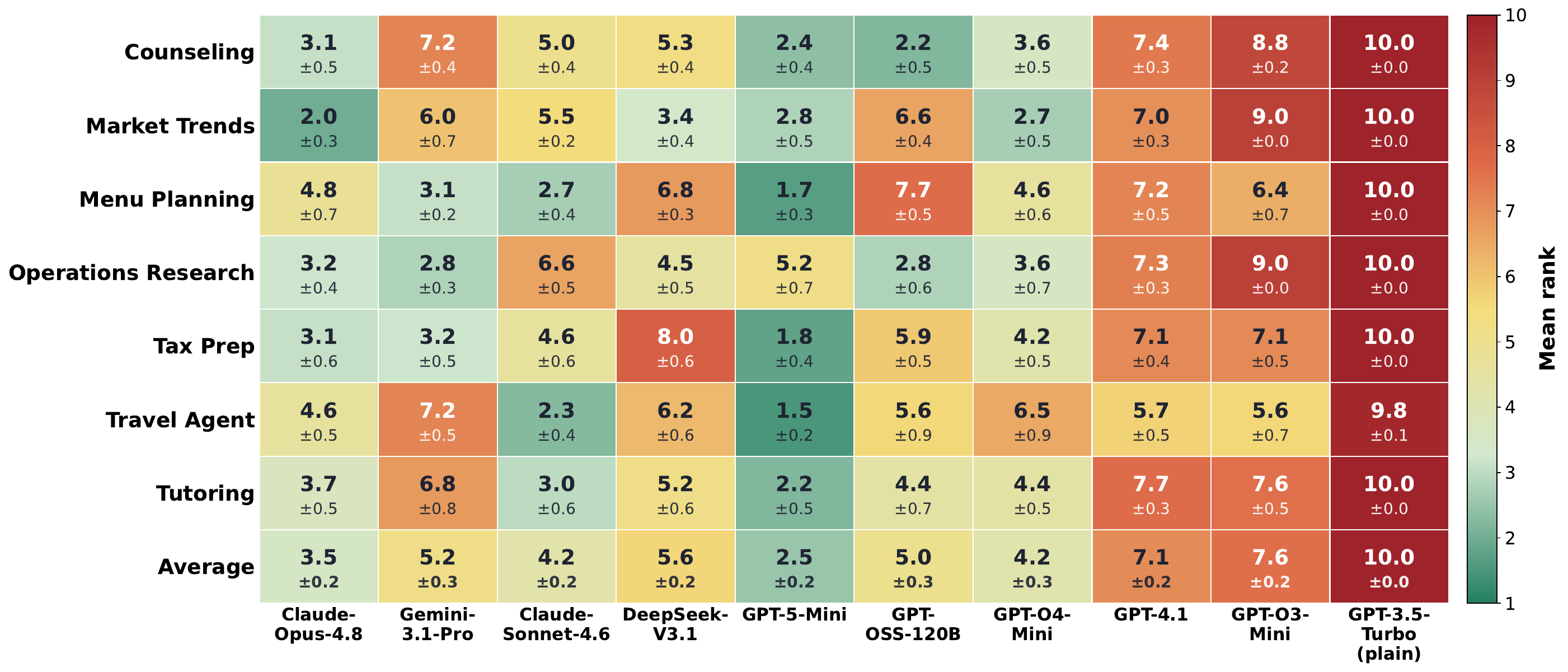}
    \caption{Automation mean ranks by task with standard error across ten independent runs.}
    \label{fig:automation-heatmap-appendix}
\end{figure}

\begin{figure}[H]
    \centering
    \includegraphics[width=1.02\linewidth]{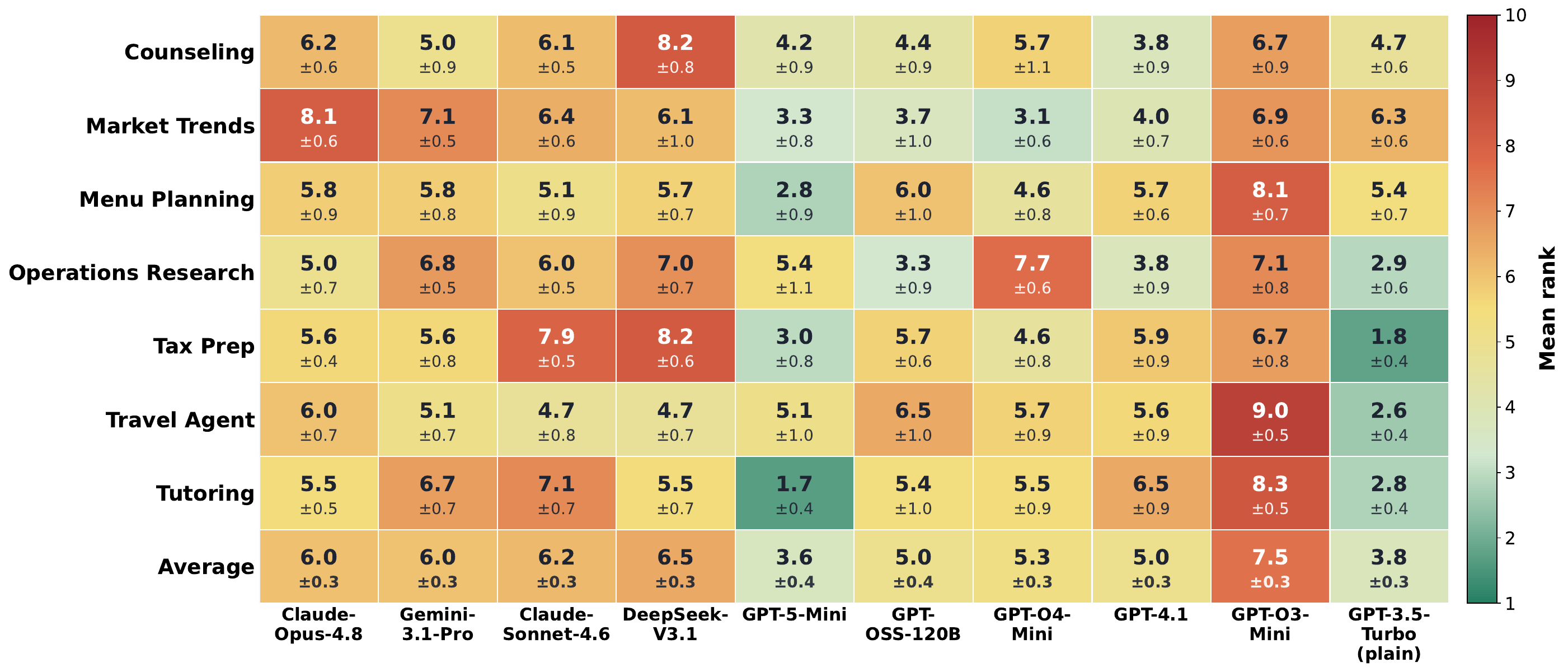}
    \caption{Augmentation mean ranks by task with standard error across ten independent runs.}
    \label{fig:augmentation-heatmap-appendix}
\end{figure}

\section{Best Model by Task and Usage Mode}
\label{app:best-model-table}

\begin{table}[H]
    \centering
    \caption{\textbf{Best-performing model by task and regime}, based on lowest mean rank across ten independent runs.}
    \label{tab:best-model-by-task}
    \small
    \begin{tabular}{lcc}
        \toprule
        Task & Automation & Augmentation \\
        \midrule
        Counseling & GPT-OSS-120B & GPT-4.1 \\
        Market Trends & Claude-Opus-4.8 & GPT-O4-Mini \\
        Menu Planning & GPT-5-Mini & GPT-5-Mini \\
        Operations Research & Gemini-3.1-Pro / GPT-OSS-120B & GPT-3.5-Turbo (plain) \\
        Tax Prep & GPT-5-Mini & GPT-3.5-Turbo (plain) \\
        Travel Agent & GPT-5-Mini & GPT-3.5-Turbo (plain) \\
        Tutoring & GPT-5-Mini & GPT-5-Mini \\
        Average (all tasks) & GPT-5-Mini & GPT-5-Mini \\
        \bottomrule
    \end{tabular}
\end{table}

\section{General Rubric Profiles}
\label{app:general-rubrics}

\begin{figure}[H]
    \centering
    \includegraphics[width=0.82\linewidth]{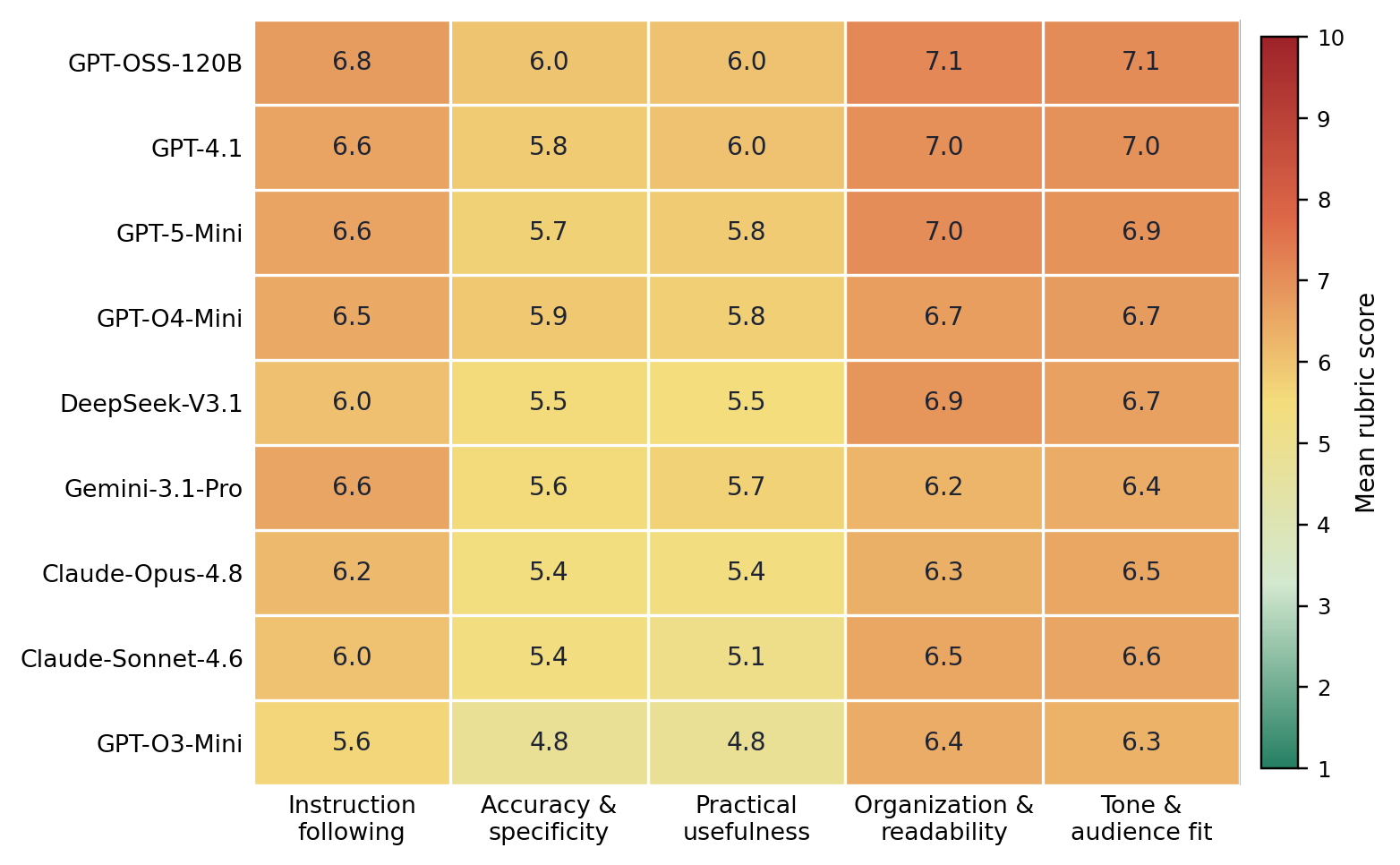}
    \caption{
    \textbf{Average general-rubric scores in the augmentation mode, aggregated across tasks and judges.}
    Scores are shown for instruction following, accuracy and specificity, practical usefulness, organization and readability, and tone/audience fit.
    Profiles are relatively flat across dimensions for most models, indicating that ranking differences are not driven by specialization on a single general rubric axis.
    }
    \label{fig:general-rubric-profiles}

\end{figure}

\section{Validity of LLM-as-Judge Evaluation}
\label{app:llm-judge-validity}

A limitation of this study is that final evaluations are conducted by LLM judges rather than by domain experts or crowdworkers. We use LLM judges because the framework is intended as a scalable pilot-stage evaluation pipeline across many models, tasks, and usage modes. We use multiple judges and leave-family-out restrictions to reduce self-preference and model-family bias. A model is not used to judge outputs from its own model family. We then aggregate rankings across eligible judges. This does not eliminate all judge bias, but it reduces the most direct form of self-evaluation and allows judge-level disagreement to be inspected. Therefore, we do not treat LLM judgments as ground truth. Instead, we implement several checks designed to assess whether the judging procedure is internally coherent and directionally valid.

First, all judges use explicit task-specific and general rubrics rather than unconstrained preference judgments. Each pairwise comparison requires the judge to score both responses across micro-rubric dimensions before selecting a winner. This design makes the basis for each judgment auditable and allows us to compare final choices against the underlying rubric scores.

\begin{figure}[H]
    \centering
    \includegraphics[width=\linewidth]{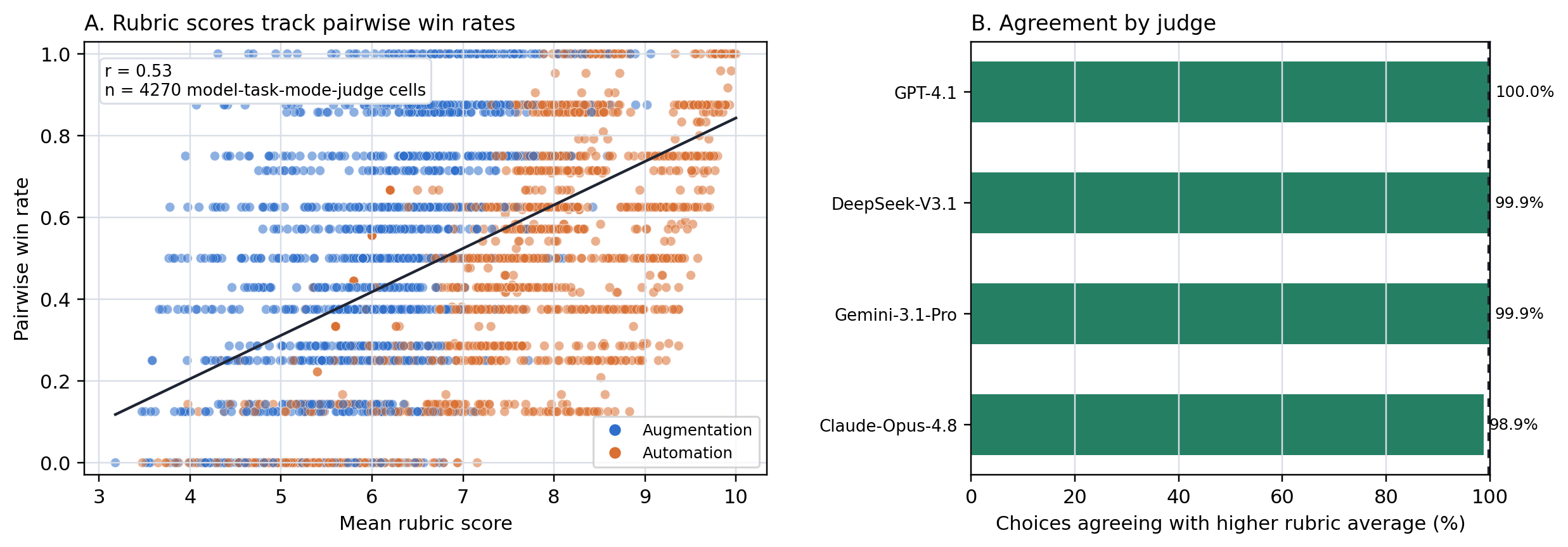}
    \caption{
    \textbf{Validation of the rubric-guided pairwise judging procedure across ten independent runs.} Panel A compares each model--task--mode--judge cell's mean rubric score with its pairwise win rate ($r=0.53$; $n=4{,}200$ cells). Panel B reports, for each judge, the percentage of non-tied comparisons in which the final pairwise choice agrees with the option receiving the higher average rubric score. Agreement ranges from 98.9\% to 100.0\% across judges.}
    \label{fig:pairwise-rubric-validation}
\end{figure}

Second, we test whether pairwise choices align with rubric scores. Across 15,120 non-tied scored comparisons, judges select the option with the higher average rubric score in 99.7\% of cases. Choice–score consistency ranges from 98.9\% to 100.0\% across the four individual judges. At the model--task--mode--judge level, mean rubric scores are positively correlated with pairwise win rates ($r=0.53$ across 4,200 cells). This indicates that the pairwise tournament rankings are directionally consistent with the underlying dimension-level evaluations.

Third, we separately assess inter-judge reliability on output comparisons evaluated by at least two eligible judges. The leave-family-out design creates a partially crossed panel because a judge is ineligible whenever a comparison contains an output from its own model
family. We consequently report both raw agreement percentage and Krippendorff's nominal alpha, which accommodates missing judge and adjusts for chance agreement. Before calculation, we map each choice to a canonical output-pair order so that randomized presentation as Option~1 or Option~2 cannot affect agreement. We retain one evaluation per judge and comparison, preventing comparisons with repeated evaluations from receiving additional weight.

\begin{figure}[H]
    \centering
    \includegraphics[width=\linewidth]{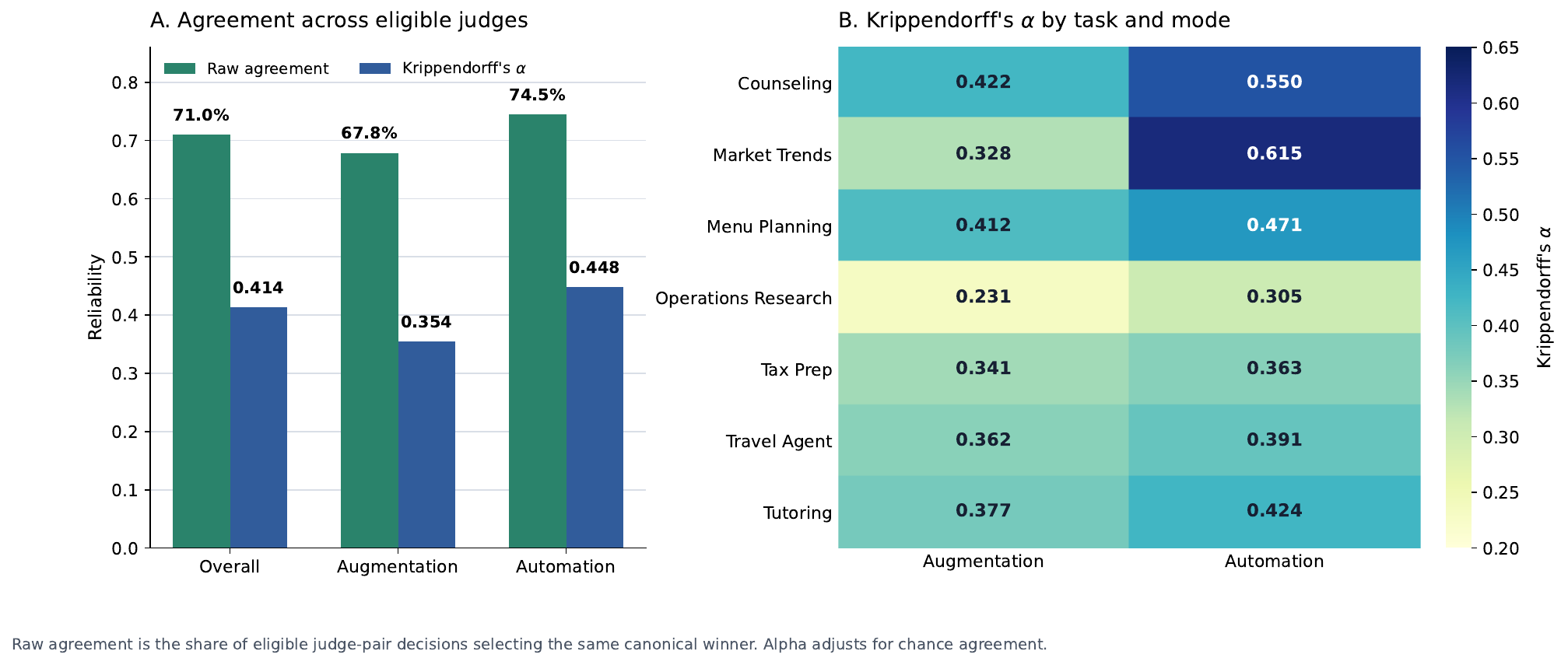}
    \caption{
    \textbf{Inter-judge reliability across ten independent runs.} Panel A reports raw agreement and Krippendorff's nominal alpha overall and separately for augmentation and automation. Raw agreement is the share of
    eligible judge-pair decisions selecting the same winner; alpha adjusts for chance agreement and accommodates missing ratings induced by leave-family-out evaluation. Panel B reports alpha by task and usage mode.
    Only comparisons evaluated by at least two eligible judges are included.}
    \label{fig:interjudge-reliability}
\end{figure}

Across 6,265 output comparisons evaluated by at least two eligible judges, the panel yields 11,257 judge-pair decisions. Judges select the same winner
in 71.0\% of these decisions, with Krippendorff's nominal
$\alpha=0.414$. Reliability is higher in automation
(74.5\%; $\alpha=0.448$) than in augmentation
(67.8\%; $\alpha=0.354$). Thirty-five comparisons with fewer than two valid ratings are excluded from the reliability calculation.

Task-mode reliability varies substantively. Krippendorff's alpha ranges from 0.231 for operations research augmentation to 0.615 for market-trends automation. Automation produces higher agreement than augmentation overall and in every task shown in Figure~\ref{fig:interjudge-reliability}. These
results indicate a meaningful but moderate common signal across judges rather than near-consensus. The lower agreement in augmentation also suggests that evaluating the downstream effects of assistance texts is more judge-dependent than evaluating direct task outputs. This finding further motivates the human-expert validation proposed in the limitations section.

\begin{figure}[H]
    \centering
    \includegraphics[width=\linewidth]{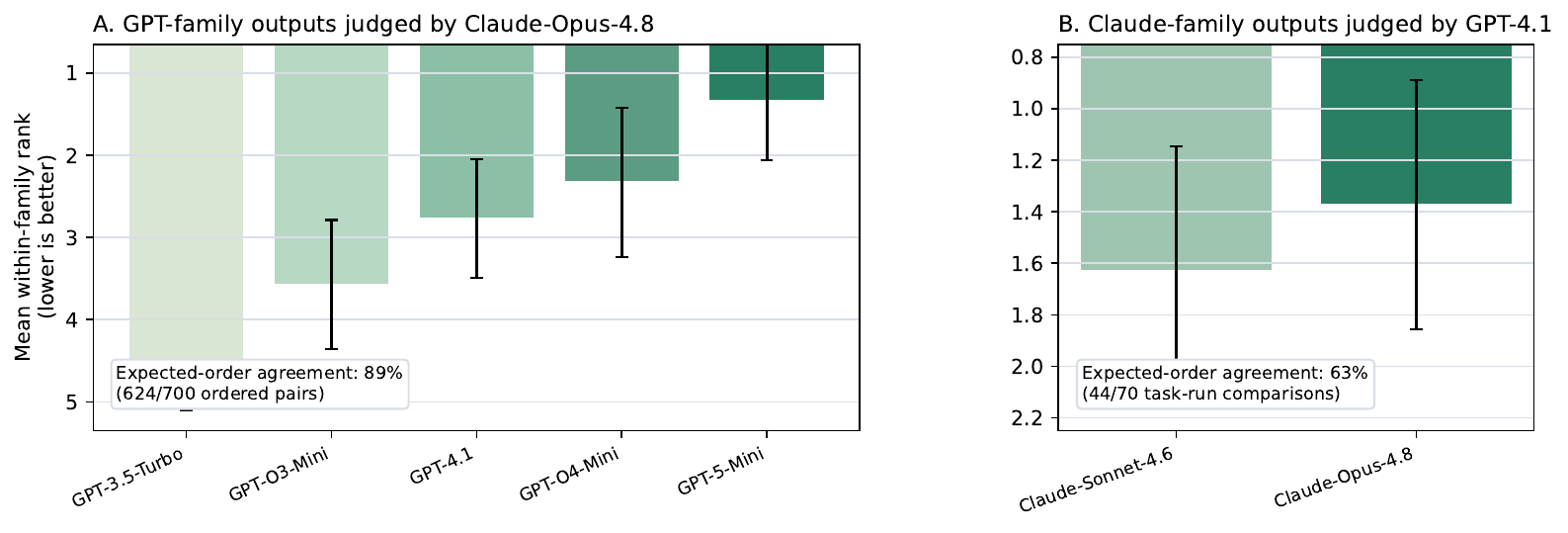}
    \caption{\textbf{Criterion-validity checks using automation outputs across ten independent runs.} Panel A reports mean within-family rank for GPT-family models judged by Claude-Opus-4.8; the judge recovers the prespecified release-order gradient in 89\% of ordered comparisons. Panel B reports mean within-family rank for Claude-family models judged by GPT-4.1; Claude-Opus-4.8 is preferred in 62.9\% of task--run comparisons. Error bars show one standard deviation across task--run cells. Lower ranks indicate better performance.
    }
    \label{fig:criterion-validity}
\end{figure}

Fourth, we conduct criterion-validity checks using automation outputs, where external expectations about model capability are clearer. Figure~\ref{fig:criterion-validity} reports two held-out-judge checks. First, Claude-Opus-4.8 evaluates GPT-family automation outputs from GPT-3.5-Turbo, GPT-O3-Mini, GPT-4.1, GPT-O4-Mini, and GPT-5-Mini. Across ten runs and seven tasks, it recovers the prespecified release-order gradient in 89\% of ordered model-pair comparisons (624 of 700), with GPT-5-Mini receiving the best mean within-family rank. Second, GPT-4.1 evaluates Claude-Sonnet-4.6 and Claude-Opus-4.8. GPT-4.1 prefers Claude-Opus-4.8 in 62.9\% of the 70 task--run
comparisons, and Opus receives the better mean within-family rank (1.37 versus 1.63). These checks provide an internal directional validation that the judging procedure recovers the prespecified within-family gradients used in this exercise. They, however, do not imply that the complete automation ordering should
reproduce external benchmark rankings.

Finally, we preserve the full evaluation trace: rompts, assistance texts, final outputs, pairwise choices, rubric scores, and judge rationales. This makes the results inspectable and readers can examine whether rankings are supported by the underlying judgments and whether specific task-level outcomes appear substantively reasonable. It is important to stress that even with such robustness checks, they do not substitute for human expert validation. The strongest future validation would compare LLM judge rankings against domain expert rankings for a subset of tasks, especially tax preparation, counseling, tutoring, and operations research. We therefore interpret the current results as a scalable simulation-based benchmark and hypothesis-generation tool, and not as a final human-validated measure of workplace value.

\section{Outputs for GPT-4.1 and DeepSeek-V3.1}
\label{app:eval-examples}

\FloatBarrier
\begin{tcolorbox}[outputbox gpt={GPT-4.1 (Rank 1 augmentation) Run 1}]
\small
Dear [Client],

I want to first acknowledge how challenging it must be to feel emotionally drained, anxious about work, and uncertain about the direction of your life. It's completely valid to feel this way, and you're not alone in experiencing these overwhelming emotions.

As we explore what you're going through, it's important to consider potential patterns that might be contributing to your current state. It could be that burnout, perfectionism, avoidance of difficult emotions, low self-efficacy, or feeling disconnected from your values are playing a role. However, please know that these are just possibilities to consider, and not labels or diagnoses.

In navigating these feelings, evidence-based approaches like Cognitive Behavioral Therapy (CBT), motivational interviewing, and positive psychology can offer valuable insights. CBT can help you reframe negative thought patterns and develop healthier coping strategies. Motivational interviewing can guide you in exploring your values and goals, while positive psychology can help you focus on your strengths and sources of resilience.

Moving forward, I encourage you to take concrete steps to manage stress, reframe unhelpful thoughts, clarify your values, and set achievable goals. This might involve practicing mindfulness or relaxation techniques, challenging negative self-talk, reflecting on what truly matters to you, and setting small, manageable objectives to work towards.

If at any point you feel overwhelmed or find it difficult to cope, please remember that it's okay to seek professional support. Your well-being is important, and reaching out for help is a sign of strength, not weakness.

As we conclude, I want to remind you that you have the power to shape your future and create a life that aligns with your values and aspirations. You are capable of overcoming these challenges and finding a sense of purpose and fulfillment. Stay hopeful and take one step at a time towards a brighter tomorrow.

Take care, and remember that I'm here to support you on this journey.

\medskip
Warm regards,\\
{[Your Name]}
\end{tcolorbox}

\newpage

\begin{tcolorbox}[outputbox ds={DeepSeek-V3.1 (Rank 10 augmentation) Run 1}]
\small
Dear [Client],

I want to first acknowledge the weight of the emotions you're carrying right now. Feeling emotionally drained, anxious about work, and uncertain about life's direction can be incredibly overwhelming. It's important to recognize and honor these feelings without downplaying their impact on you.

In exploring your current state, it's possible that patterns like burnout, perfectionism, avoidance, low self-efficacy, or a mismatch in values might be at play. While I'm not here to diagnose, understanding these patterns can shed light on the challenges you're facing.

Drawing from evidence-informed approaches like Cognitive Behavioral Therapy (CBT), Motivational Interviewing, and Positive Psychology, we can work together to reframe your thoughts, regulate stress, clarify your values, and set achievable goals. By breaking down these overwhelming feelings into manageable steps, we can pave a path towards a more fulfilling and balanced life.

It's essential to prioritize your well-being, so if the distress you're experiencing feels severe or persistent, seeking professional support can provide additional guidance and resources. Your mental health is paramount, and reaching out for help is a courageous step towards healing.

As we navigate this journey together, remember that you hold the power to shape your future. Your agency and resilience are remarkable strengths that will guide you towards a brighter tomorrow. Stay hopeful, stay strong, and know that I'm here to support you every step of the way.

\medskip
Warm regards,\\
{[Your Name]}
\end{tcolorbox}

\FloatBarrier

\end{document}